\documentclass[conference]{IEEEtran}
\IEEEoverridecommandlockouts
\usepackage{cite}
\usepackage{amsmath}
\usepackage{algorithmic}
\usepackage{graphicx}
\usepackage{textcomp}
\usepackage{xcolor}
\usepackage{tcolorbox}
\tcbuselibrary{breakable}
\def\BibTeX{{\rm B\kern-.05em{\sc i\kern-.025em b}\kern-.08em
    T\kern-.1667em\lower.7ex\hbox{E}\kern-.125emX}}

\usepackage{graphicx}
\usepackage{balance}  
\usepackage{xcolor}
\usepackage{url}
\usepackage{enumitem} 
\usepackage{pifont}

\usepackage{marginnote} %

\usepackage{multirow}
\usepackage[normalem]{ulem}
\usepackage{caption}
\usepackage{subcaption}

\usepackage{listings}
\usepackage{bm}
\usepackage[linesnumbered,ruled,vlined]{algorithm2e}
\usepackage{graphicx}
\usepackage{hyperref}

\usepackage{makecell}
\usepackage{fontawesome}

\usepackage{multicol}

\usepackage{balance} 
\usepackage{booktabs} 
\usepackage{caption}
\usepackage{comment}
\usepackage{enumitem}
\usepackage{url}
\setlist[itemize]{leftmargin=2.2em}

\usepackage{fancyvrb} 
\usepackage[symbol]{footmisc}
\usepackage{graphicx} 
\usepackage{makecell} 
\usepackage{multirow} 
\usepackage{soul}
\usepackage{subcaption}
\usepackage{xcolor}
\usepackage{xspace} 
\usepackage{bm}
\usepackage{utfsym}
\usepackage{cancel}

\usepackage{graphicx}
\usepackage{balance}  
\usepackage{xcolor}
\usepackage{url}
\usepackage{enumitem} 
\usepackage{pifont}

\usepackage{multirow}
\usepackage[normalem]{ulem}
\usepackage{caption}
\usepackage{subcaption}

\usepackage{listings}
\usepackage{bm}
\usepackage[linesnumbered,ruled,vlined]{algorithm2e}
\usepackage{graphicx}
\usepackage{setspace}

\usepackage{makecell}
\usepackage{fontawesome}

\usepackage{multicol}

\usepackage{amsmath} 
\usepackage{balance} 
\usepackage{booktabs} 
\usepackage{caption}
\usepackage{comment}
\usepackage{enumitem}
\setlist[itemize]{leftmargin=2.2em}

\usepackage{fancyvrb} 
\usepackage[symbol]{footmisc}
\usepackage{graphicx} 
\usepackage{makecell} 
\usepackage{multirow} 
\usepackage{soul}
\usepackage{subcaption}
\usepackage{xcolor}
\usepackage{xspace} 
\usepackage{bm}
\usepackage{utfsym}
\usepackage{cancel}
\usepackage{authblk}
\usepackage{marginnote}

\SetKwInOut{Input}{input}\SetKwInOut{Output}{output}

\usepackage{color}
\usepackage{xcolor}
\usepackage{pifont}
\usepackage{float}

\newcommand{\whiteding}[1]{\ding{\numexpr171+#1\relax}}
\newcommand{\blackding}[1]{\ding{\numexpr181+#1\relax}}

\newcommand\dbname{\ensuremath{\texttt{GeoTP}}\xspace}
\newif\ifextended\extendedfalse

\newcommand{\maintext}[1]{\ifextended\relax\else#1\fi} 

\newcommand{\extended}[1]{\ifextended#1\else\relax\fi}

\newcommand\firstlocktime{\ensuremath{\hat{t}_{1st}^{T_{ij}}}}
\newcommand\lastunlocktime{\ensuremath{\check{t}_{last}^{T_{ij}}}}
\newcommand\transactionstarttime{\ensuremath{t_{start}^{T_{ij}}}}

\begin{document}

\extendedtrue

\maintext{\title{\dbname: Latency-aware Geo-Distributed Transaction Processing in Database Middlewares}}
\extended{\title{\dbname: Latency-aware Geo-Distributed Transaction Processing in Database Middlewares (Extended Version)}}

\author{
 Qiyu Zhuang$^\dagger$, Xinyue Shi$^\dagger$, Shuang Liu$^\dagger$, Wei Lu$^\dagger$, Zhanhao Zhao$^\dagger$\\ Yuxing Chen$^\ddagger$, Tong Li$^\dagger$, Anqun Pan$^\ddagger$, Xiaoyong Du$^\dagger$\\

\vspace{1.8mm}
\fontsize{10}{10}\textit{$^\dagger$ Renmin University of China}
    \qquad\fontsize{10}{10}\textit{$^\ddagger$ Tencent Inc.} 
 \\\fontsize{9}{9}\texttt{$^\dagger$\{qyzhuang, xinyueshi, shuang.liu, lu-wei, zhanhaozhao, tong.li, duyong\}@ruc.edu.cn} 
 \\\fontsize{9}{9}\texttt{$^\ddagger$\{axingguchen, aaronpan\}@tencent.com}  
}
\maketitle
\thispagestyle{plain}
\pagestyle{plain}

\begin{abstract}

The widespread adoption of database middleware for supporting distributed transaction processing is prevalent in numerous applications, with heterogeneous data sources deployed across national and international boundaries. 
However, transaction processing performance significantly drops due to the high network latency between the middleware and data sources and the long lock contention span, where transactions may be blocked while waiting for the locks held by concurrent transactions.
In this paper, we propose \dbname, a latency-aware geo-distributed transaction processing approach in database middleware. 
\dbname incorporates three key techniques to enhance performance in geo-distributed scenarios.
First, we propose a decentralized prepare mechanism to reduce network round-trips for distributed transactions. 
Second, we design a latency-aware scheduler to minimize the lock contention span by strategically delaying the lock acquisition.
Third, heuristic optimizations are proposed for the scheduler to reduce the lock contention span further.
We implemented \dbname on Apache Shardingsphere, a state-of-the-art middleware, and extended it into Apache ScalarDB. 
Experimental results on YCSB and TPC-C demonstrate that \dbname achieves up to 17.7x performance improvement. 
\end{abstract}

\setcounter{page}{1}
\setcounter{section}{0}
\setcounter{figure}{0}
\setcounter{table}{0}
\section{Introduction}
\label{intro}
Globalization of enterprises is an inevitable trend in economic development.
Critical global applications, such as cross-border e-commerce~\cite{e-commerce} and e-banking~\cite{e-banking}, require data to be stored 
in different regions for compliance with local government regulations~\cite{tiktok-data-store}, while guaranteeing atomic transaction processing among them.
For instance, a global e-commerce application might store its US user account data in the US and the stock data in the warehouse location (Singapore). 
Furthermore, these databases are often managed by different departments, making them highly likely to be heterogeneous. 
Consequently, a typical product purchase requires a geo-distributed transaction that simultaneously updates two heterogeneous databases in different locations, ensuring the user balance and current stock are updated atomically.
To achieve this, database middlewares, such as Shardingsphere~\cite{shardingsphere} and ScalarDB~\cite{DBLP:journals/pvldb/YamadaSIN23}, become indispensable to connect heterogeneous databases across different regions for unified data services. 
Unlike distributed database systems~\cite{DBLP:journals/tocs/CorbettDEFFFGGHHHKKLLMMNQRRSSTWW13,yugabyte,DBLP:conf/sigmod/TaftSMVLGNWBPBR20,DBLP:journals/pvldb/ChenPLYHTLCZD24_TDSQL} 
, which usually requires rebuilding databases and applications. Database middleware can provide transaction processing capabilities without modification. 
This facilitates easier global service constructions, leading to widespread adoption in enterprises~\cite{sharding_application, salesforce-for-Neo4j}.

\begin{figure}[t]
    \centering
\begin{subfigure}{0.46\linewidth}
\includegraphics[width=\linewidth]{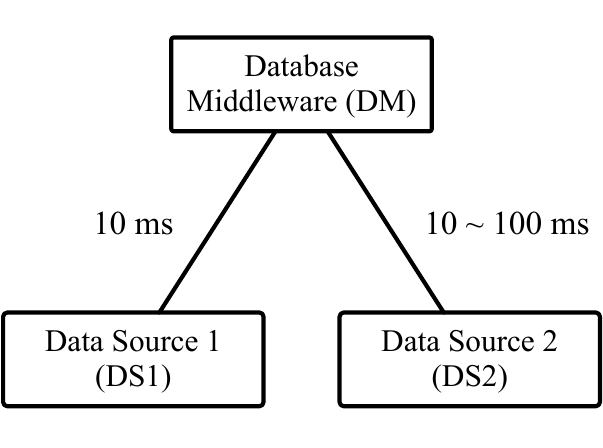}
\vspace{-5mm}
\caption{Data source deployment}
\label{intro.deploy}
\end{subfigure}
\begin{subfigure}{0.46\linewidth}
\includegraphics[width=\linewidth]{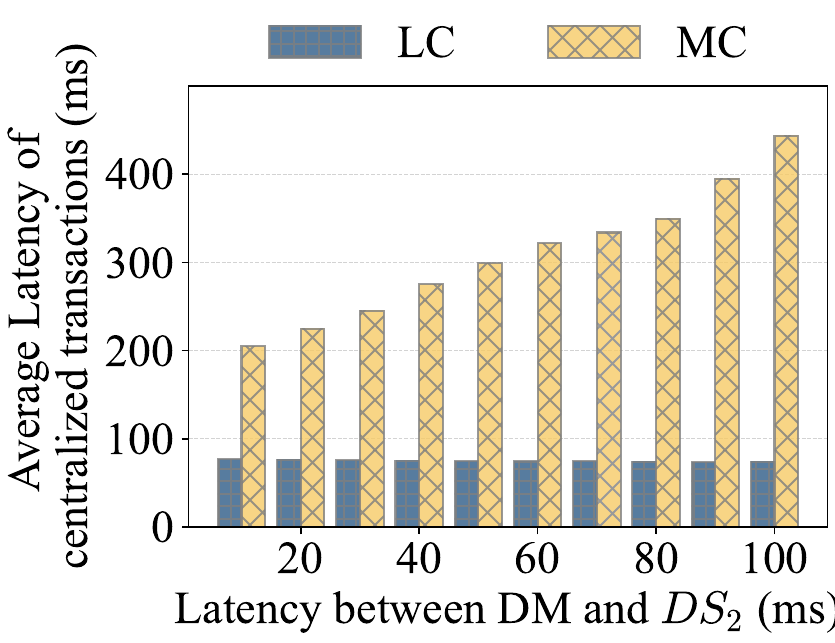}
\vspace{-5mm}
\caption{Impact of network latency}
\label{intro.cdf}
\end{subfigure}
\vspace{-2mm}
\caption{A motivating example
}
\vspace{-6mm}
\label{intro.motivation}
\end{figure}


Database middlewares (abbreviated as DMs) typically employ the eXtended Architecture (XA) Protocol, an extension of the two-phase commit (2PC), to ensure the transaction's atomicity.
Databases, such as MySQL~\cite{mysql} and PostgreSQL~\cite{postgresql}, serve as the data sources of DMs. 
Specifically, the DM accepts the transactions submitted by the clients.
We consider transaction $T$ as a \textit{centralized transaction} if it involves a single data source; otherwise, $T$ is considered as a \textit{distributed transaction}.
For a {\it centralized transaction}, the DM forwards it to the relevant data source, which executes it and returns the results. Upon receiving the commit or abort command, the DM instructs the relevant data source to commit or abort directly, requiring one wide-area network (WAN) round trip.  
For a {\it distributed transaction},
whether interactive or stored procedures, the typical transaction processing protocol first executes read/write operations in the relevant data sources during the execution phase. Upon receiving the commit or abort command, the DM follows the 2PC~\cite{DBLP:journals/pvldb/MaiyyaNAA19, DBLP:conf/sosp/ZhangSSKP15}, including a prepare phase and a commit phase, to ensure transaction atomicity. This commit process requires two WAN round trips.
\textit{The WAN round trip time dominates transaction latency and significantly degrades performance, particularly for {\it distributed transactions}, which requires two WAN round trips for commitment. The impact is more pronounced in geo-distributed scenarios where network latencies between the DM and data sources are high}~\cite{DBLP:journals/pvldb/ZhangLZXLXHYD23}. 
Data sources \cite{mysql,DBLP:conf/sigmod/AntonopoulosBDS19,postgresql,DBLP:journals/pvldb/BarthelsMTAH19,DBLP:journals/pvldb/WuALXP17} typically use two-phase locking (2PL) or its variants for concurrency control\footnote{We focus on serializable isolation level in this work.}. \textit{In addition to the overhead of WAN communications, long lock contention span—the time span between the acquisition (before reads or writes) and release (after the commitment) of a record lock—is also a critical factor for performance degradation.} 
We explicitly design an experiment to show the impact of lock contention span on transaction performance. 
Figure~\ref{intro.motivation} illustrates two data sources, $DS_1$ and $DS_2$. with a 10 ms WAN round-trip time (RTT) between the DM and $DS_1$, and varying latency between the DM and $DS_2$ (10–100 ms). The workload includes 80\% {\it centralized transactions} accessing $DS_1$ and 20\% {\it distributed transactions} accessing both $DS_1$ and $DS_2$. 
We evaluate the average latency of {\it centralized transactions} (on $DS_1$) with varying the network latency between DM and $DS_2$ under low-contention (LC) and medium-contention (MC) workloads. 
As depicted in Figure~\ref{intro.cdf}, 
network latency between the DM and $DS_2$ has a more substantial impact on {\it centralized transactions} under medium contention than low contention, even though these transactions do not access $DS_2$.
This is because in medium-contention workloads, {\it centralized transactions} are more likely to access shared records with {\it distributed transactions} that access $DS_2$. 
The lock contention span of {\it distributed transactions}, significantly affected by the network latency between the DM and $DS_2$, impacts the latency of {\it centralized transactions} due to shared record blocking. 
We provide more details for this in \S\ref{sec.back1}. 

Several works are proposed to reduce WAN round trips in distributed transaction processing.
Early Prepare~\cite{DBLP:conf/srds/StamosC90} and RedT~\cite{DBLP:journals/pvldb/ZhangLZXLXHYD23} reduce the network round-trips by writing logs during execution, thus eliminating the prepare phase. 
Carousel~\cite{DBLP:conf/sigmod/YanYZLWSB18}, Natto~\cite{DBLP:conf/sigmod/YangY022} and Janus~\cite{DBLP:conf/osdi/MuNLL16} reduce network round trips by integrating consensus protocols with 2PC, assuming knowing the read/write sets in advance.  
However, they require rewriting the kernel-level protocol, making them difficult to extend to heterogeneous data sources. 
Another line of work has proposed delayed scheduling techniques to reduce lock contention spans. 
QURO~\cite{DBLP:journals/pvldb/YanC16} preprocesses the application code to reorder the read/write operations and delays the acquisition of exclusive locks for writes. However, it lacks consideration for network latency, limiting its effectiveness in geo-distributed scenarios.
Chiller~\cite{DBLP:conf/sigmod/ZamanianSBK20} and DAST \cite{DBLP:conf/eurosys/ChenSJRLWZCC21} address latency differences in geo-distributed scenarios by scheduling cross-region subtransactions to follow intra-region ones, as hot records are often in the intra-region ones. However, these methods are designed for stored procedures and overlook the varied latency between cross-region nodes and execution times, potentially differing by orders of magnitude, leaving substantial room for optimizing the lock contention span. 
In this paper, we present \dbname, a latency-aware geo-distributed transaction processing approach in database middleware. We propose three key techniques to mitigate the impact of network latency and lock contention while ensuring that \dbname continues to support general-purpose transactions, including interactive and stored procedures. Our key techniques and contributions are summarized as follows. 

\textbf{(1) Decentralized prepare mechanism that offloads the coordination cost required for the prepare phase (\S\ref{design-1}).} 
\dbname triggers the prepare phase implicitly at the end of the execution phase, effectively eliminating one WAN round trip and reducing the latency of distributed transactions. However, this process is challenging due to the different transaction protocols used by various data sources. 
To address this, we leverage annotations to mark the last statement and develop an efficient component called geo-agent to abstract differences between data sources, facilitating decentralized preparation in \dbname. 
Additionally, we design an early abort mechanism that allows fault transactions to abort quickly, preventing such transactions from degrading the performance. 



\begin{figure}[t]
    \centering
    \includegraphics[width=0.47\textwidth]{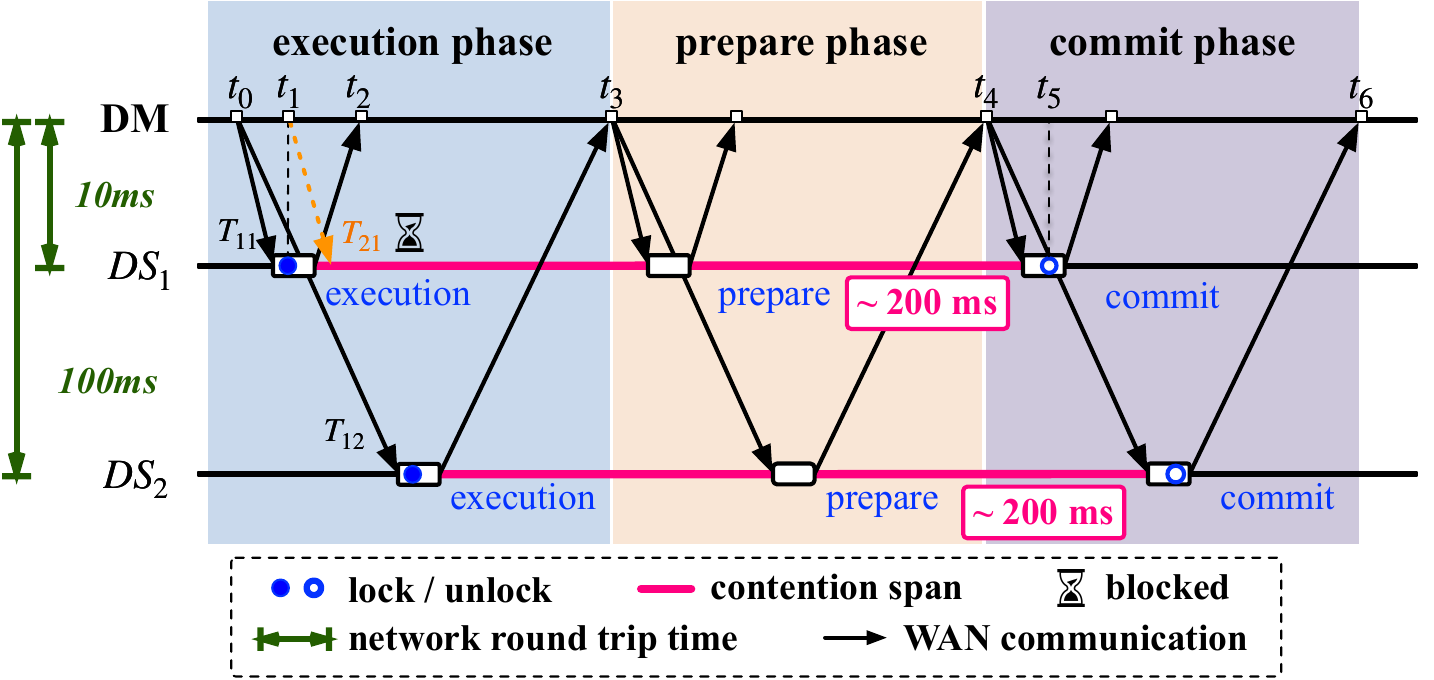}
    \vspace{-1mm}
    \caption{Distributed transaction processing in DMs}
    \label{back.xa-2pc}
    \vspace{-6mm}
\end{figure}

\textbf{(2) Latency-aware scheduling to minimize the lock contention span (\S\ref{design-2}).} 
The lock contention span of a transaction is determined by the highest network latency involved, resulting in unnecessary lock contention. 
To address this, we propose a latency-aware scheduling mechanism that postpones the lock request time point for the subtransactions accessing data sources with lower network latency. Since the lock release time point remains unchanged, the lock contention span of these subtransactions is reduced. This approach minimizes the impact of {\it distributed transactions} on transaction concurrency, thereby improving the overall system performance. 

\begin{figure*}[t]
    \centering
    \includegraphics[width=0.95\textwidth]{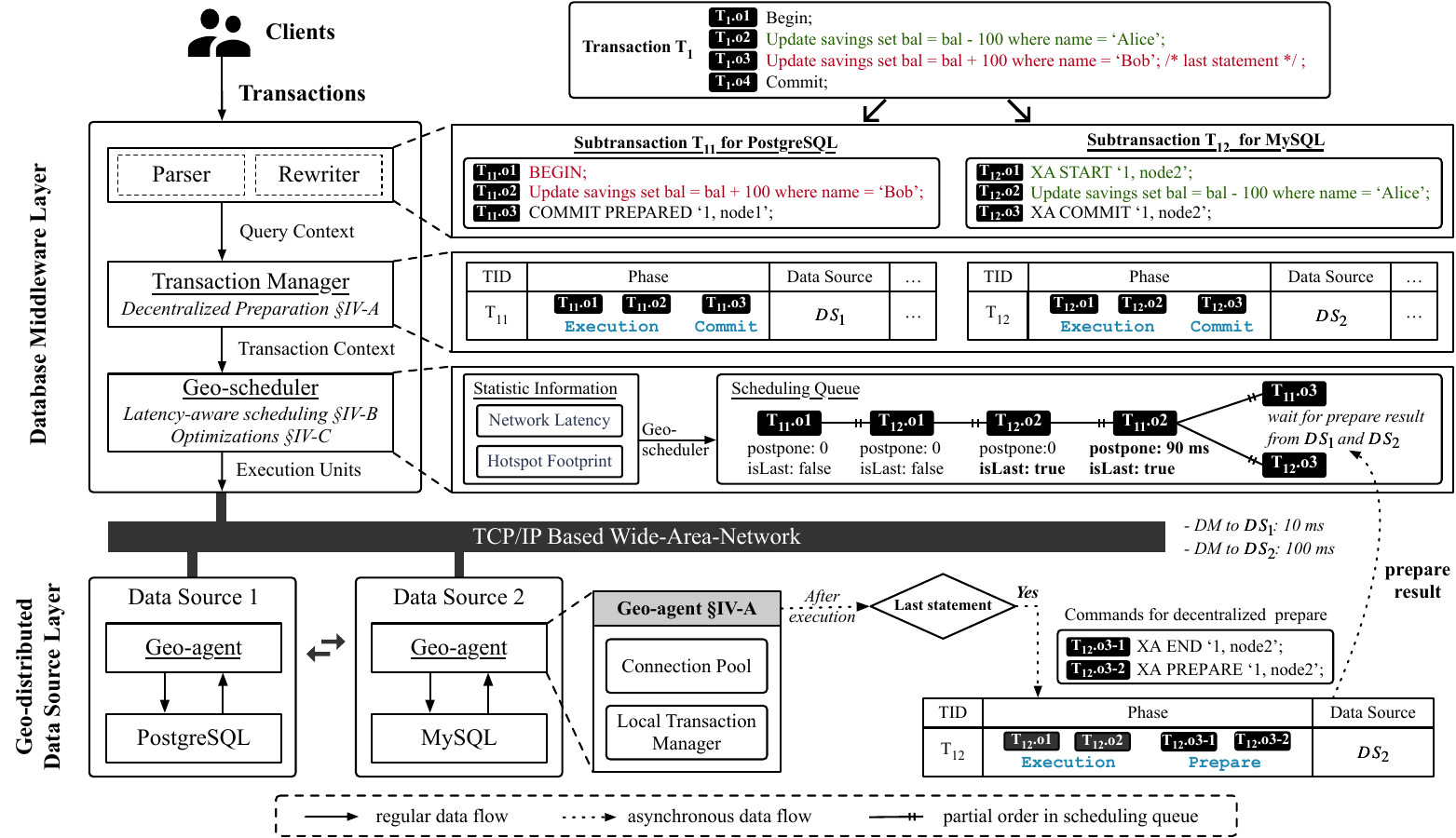}
    \vspace{-1mm}
    \caption{An overview of \dbname
    } 
    \label{over.sec1}
    \vspace{-6mm}
\end{figure*}

\textbf{(3) Optimized scheduling for high-contention workloads considering local execution latency (\S\ref{design-3}).} 
In high-contention workloads, the lock contention span is influenced not only by the highest network latency but also by the time subtransactions spend waiting to acquire locks. For instance, a subtransaction with lower network latency might still face significant latency if it has to wait a long time for locks on hotspots, causing its local execution latency to exceed the longest network latency and become a bottleneck. To enhance scheduling precision in such scenarios, we employ heuristic optimization, including transaction admission and local execution latency forecasting mechanisms. By doing this, \dbname can further reduce the lock contention span.


We implement \dbname on Apache ShardingSphere and extend our optimizations on ScalarDB. 
Extensive evaluations on YCSB and TPC-C show that \dbname achieves a performance improvement of up to 17.7x over Shardingsphere and up to 3.2x over ScalarDB and offers comparable performance to distributed databases.

\section{Motivation Example\label{sec.back1}}

In this section, we use Figure~\ref{back.xa-2pc} as an example to motivate our work. The network latency between DM and $DS_1$ is 10 ms, while the latency between DM and $DS_2$ is 100 ms. There are two transactions, $T_1$ and $T_2$, arriving DM at times $t_0$ and $t_1$, respectively. $T_1$ is a distributed transaction accessing records in $DS_1$ and $DS_2$, while $T_2$ is a centralized transaction accessing a shared record $r$ with $T_1$ in $DS_1$. We denote the subtransaction of $T_i$ executed on data source $DS_j$ as $T_{ij}$. Note that in most scenarios, network latency outweighs execution latency. For simplicity, we ignore the local execution time required in each phase without loss of generality. 


In a typical distributed transaction scenario, the DM acts as a \textit{coordinator}, while each data source serves as a \textit{participant}. The lifecycle of a distributed transaction, e.g., $T_1$, can be divided into three phases: 1) the execution phase, 2) the prepare phase, and 3) the commit phase. 
During the execution phase, the DM parses a transaction $T_1$ into subtransactions $T_{11}$ and $T_{12}$ and dispatches them to data sources $DS_1$ and $DS_2$ based on data distribution. The data source, e.g., $DS_1$, then initiates a subtransaction $T_{11}$, acquires locks on record $r$ before reads or writes, and sends the execution results back to the DM. 
The client submits the \textit{commit} request at $t_3$, triggering the prepare phase. The DM notifies $DS_1$ and $DS_2$ to verify whether the subtransactions are ready for the commitment. In response, data sources persist the transaction states and write-ahead logs and then return the prepared result. The DM collects all prepared results at $t_4$ and determines whether to commit or abort the transaction based on the return results from the data sources. 
Finally, in the commit phase, the DM dispatches the final decision to $DS_1$ and $DS_2$, which involves another WAN round trip, and the transaction is completed at $t_6$. 
The lifecycle of $T_1$ is from $t_0$ to $t_6$, involving three WAN round trips, which dominate the transaction latency. 

As evident from Figure~\ref{back.xa-2pc}, $T_{11}$ acquires the lock on $r$ at $t_1$ and release it at $t_5$. The lock contention span of $T_{11}$ on record $r$ is around 200 ms (2 WAN round trips), determined by the network latency between the DM and $DS_2$.
The subtransaction $T_{21}$ arrives $DS_1$ at around $t_2$ and is blocked by $T_{11}$ until $t_5$ due to its prolonged lock contention span. 
The DM has to await the execution results from $T_{21}$, which are received around $t_5$+5 ms, significantly increasing the execution latency of $T_2$. 
Even worse, if $T_{21}$ acquires locks on other records, the lock contention span can transitively block other concurrent transactions. 
Note that, even if transaction $T_2$ is a \textit{centralized transaction} without accessing any record in $DS_2$, the network latency between DM and $DS_2$ still affects the transaction latency of $T_2$ through the lock contention span of transaction $T_1$. 
This explains the experiment results in Figure~\ref{intro.cdf}.

\textbf{This motivation example highlights the substantial impacts of network latency and long lock contention spans on the transaction performance in geo-distributed scenarios}.

\section{Overview of \dbname\label{sec.overview}}



Figure~\ref{over.sec1} provides an overview of \dbname, which operates in the two-layer architecture. The first layer functions as the DM~\cite{DBLP:conf/icde/LiZPLWSWCGG22, Citus, MySQL_Cluster}, while the second layer comprises data sources that can be geographically distributed and heterogeneous; for example, $DS_1$ includes a PostgreSQL instance and $DS_2$ includes a MySQL instance. 
For clarity, 
we assign a monotonic identifier to each operation within a transaction $T$.
We assume that applications can use annotations, which are prefixes or suffixes on SQL statements, to pass certain operations hints to \dbname.
Given that SQL annotations are commonly used to guide and influence database query optimization~\cite{MySQL_Hint, PostgreSQL_Hint} manually, we consider this assumption reasonable.



\subsection{Database Middleware Layer}

In the first layer, similar to existing DMs, \dbname is equipped with the parser and rewriter, which accept transactions submitted by the clients and transform them into multiple subtransactions. 
Despite these components, \dbname is equipped with an enhanced transaction manager and a geo-scheduler, which differ from existing DMs. The enhanced transaction manager is responsible for coordinating the execution and handling failure recovery. The geo-scheduler is particularly crafted to calculate the optimal start time point for subtransactions, minimizing the lock contention span. 
Next, we use transaction $T_1$ in Figure~\ref{over.sec1} as an example to explain the transaction processing in \dbname and our key techniques.  
Suppose Alice submits transaction $T_1$, which transfers \$100 from her account to Bob's account.
Note that Bob's account is stored in a PostgreSQL instance ($DS_1$), and Alice's account is stored in a MySQL instance ($DS_2$). 

\vspace{1mm}
\noindent\textbf{Parser and rewriter.} These components parse SQL statements received from clients and then rewrite them according to the grammar rules of target databases. For example, they translate $T_1$ into ${T_{11}}$ and ${T_{12}}$, which are executable for PostgreSQL and MySQL, respectively. 
Operations ${T_{11}.o1}$ and ${T_{12}.o1}$ start an XA transaction in each data source. Operation ${T_{11}.o2}$ deposits \$100 into Bob's account in PostgreSQL, while ${T_{12}.o2}$ deducts \$100 from Alice's account in MySQL. Subsequently, ${T_{11}.o3}$ and ${T_{12}.o3}$ attempt to commit the respective subtransactions. 
\vspace{1mm}
\noindent\textbf{Transaction manager.} 
Unlike conventional transaction managers, our enhanced transaction manager employs a decentralized prepare mechanism to eliminate the prepare phase from the critical path of XA protocol (\S\ref{design-1}). In our design, the prepare phase is no longer triggered by \textit{commit} request of clients. Instead, it is initiated after the last statement in the execution phase, explicitly specified by the client (e.g., $T_1.o3$).
Upon identifying the last SQL statement, the transaction manager 
combines the decentralized prepare phase with the processing of this statement over the underlying data sources.  
For example, since there is no dependency between $T_1.o2$ and $T_1.o3$, we assume the client sends them together to the DM. The transaction manager treats them as the last SQL statement of each data source. 
Importantly, if some data sources are involved in the transaction but not processing the last SQL statement, the transaction manager directly notifies those data sources to initiate the prepare phase. 
This approach eliminates one WAN round trip required by the prepare phase for distributed transactions.



\vspace{1mm}
\noindent\textbf{Geo-scheduler.} 
The geo-scheduler implements the latency-aware scheduling of subtransactions by calculating each statement's optimal start time point based on the network latency and predicted transaction execution latency.
As is shown in Figure~\ref{over.sec1}, suppose the average network latency from the DM to $DS_1$ and $DS_2$ is 10 ms and 100 ms. 
We show the schedule produced by the geo-scheduler on the bottom-right of the first layer.
Specifically, $T_{11}.o1$, and $T_{12}.o1$ are first scheduled to execute. 
Next, $T_{12}.o2$ is scheduled without postponing, while $T_{11}.o2$ is scheduled and has been postponed 90 ms for execution.
Unlike traditional schedulers where $T_{11}.o2$ and $T_{12}.o2$ are sent to data sources simultaneously, resulting in a contention span of 100 ms for both operations; our scheduler adopts a prioritized strategy (details in \S\ref{design-2} \& \ref{design-3}). This strategy reduces the contention span of $T_{11}.o2$ to 10 ms without increasing the overall latency of $T_1$. 
The scheduling of $T_{11}.o3$ and $T_{12}.o3$ needs to wait for the prepare results from 
 data sources $DS_1$ and $DS_2$, respectively.  
\subsection{Geo-distributed Data Source Layer}
In the second layer, each data source is equipped with a geo-agent comprising two crucial components: a connection pool and a local transaction manager. 
The connection pool manages connections with the DM, the underlying database, and other geo-agents.
The local transaction manager receives/forwards messages from/to the DM or database and notifies the database to initiate the implicit decentralized prepare phase. 
Following the running example, after the execution of $T_{12}.o2$, which is the last statement of $T_{12}$, the geo-agent instructs $DS_2$ to execute $T_{12}.o3\mbox{-}1$ and $T_{12}.o3\mbox{-}2$ (shown in the bottom right corner of Figure~\ref{over.sec1}) to complete the prepare phase. Once $T_{12}.o3$ is received, the geo-agent only needs to await the result of $T_{12}.o3\mbox{-}2$ and then instructs $DS_2$ to commit $T_{12}$. 
In the event of an abort during execution, the geo-agent implements the \textit{early abort} mechanism to proactively notify other data sources involved to pre-terminate other subtransactions without the coordination of DM (\S\ref{design-1}).

\section{Detailed Design\label{sec.design}}
In this section, we introduce the key techniques of \dbname, including the decentralized prepare mechanism (\S\ref{design-1}), the latency-aware transaction scheduling mechanism (\S\ref{design-2}), and further optimizations for high-contention workloads (\S\ref{design-3}).


\subsection{Decentralized Prepare Mechanism\label{design-1}}
With the traditional 2PC protocol, the DM is responsible for coordinating both the prepare and commit phases of distributed transactions, incurring two WAN round trips and resulting in expensive coordination costs. This issue often leads to prolonged distributed transaction latency, significantly impacting performance. In \dbname, we propose (1) a decentralized prepare mechanism to eliminate one WAN round trip time for the prepare phase and (2) an early abort mechanism to pre-terminate unnecessary execution of subtransactions. 
\maintext{
\textcolor{blue}{
Due to space limitations, we illustrate the pseudo-code of key functions in our extended version \cite{GeoTP}.
}
}
\begin{figure}[t]
    \centering

    \begin{subfigure}{0.97\linewidth}
        \includegraphics[width=\linewidth]{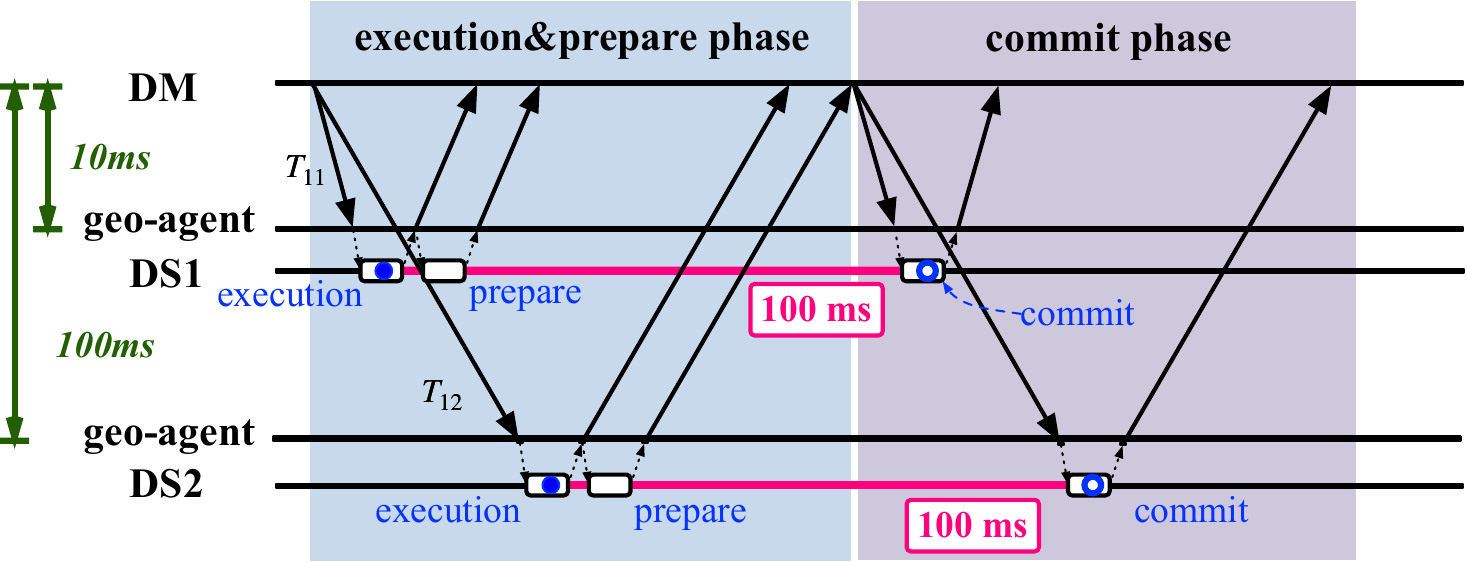}
        \vspace{-6mm}
        \caption{Decentralized prepare mechanism}
        \label{design.agent}
        \vspace{1mm}
    \end{subfigure}

    \extended{
    \begin{subfigure}{0.97\linewidth}
        \centering
        \includegraphics[width=\linewidth]{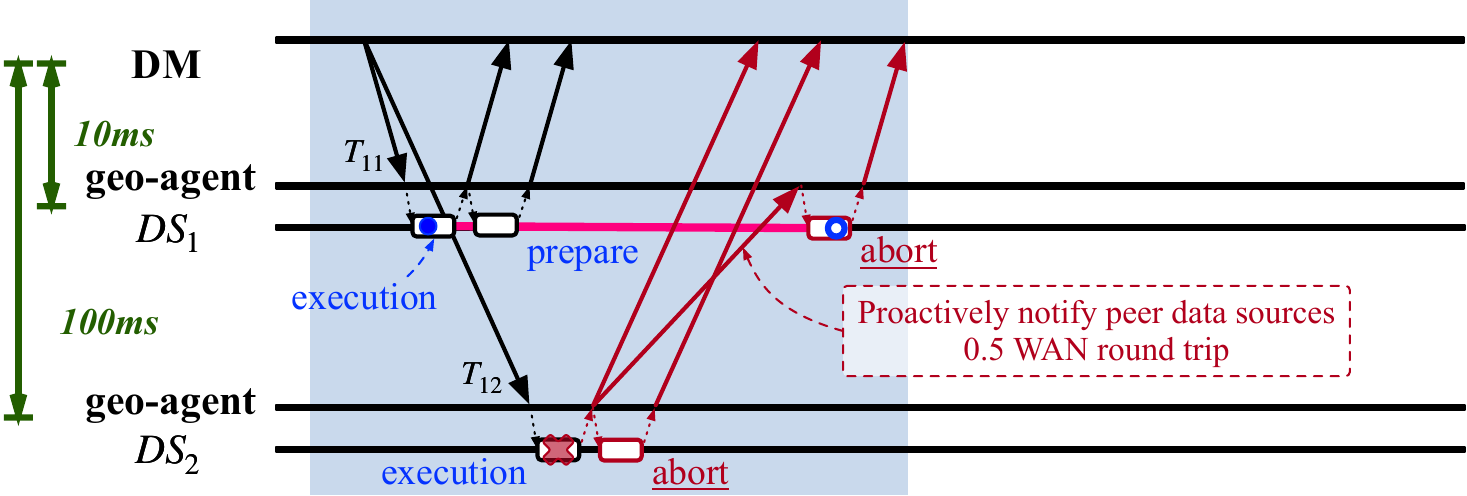}
        \vspace{-6mm}
        \caption{Early abort mechanism}
        \vspace{1mm}
        \label{desigin.rollback}
    \end{subfigure}
    }

    \begin{subfigure}{0.97\linewidth}
        \includegraphics[width=\linewidth]{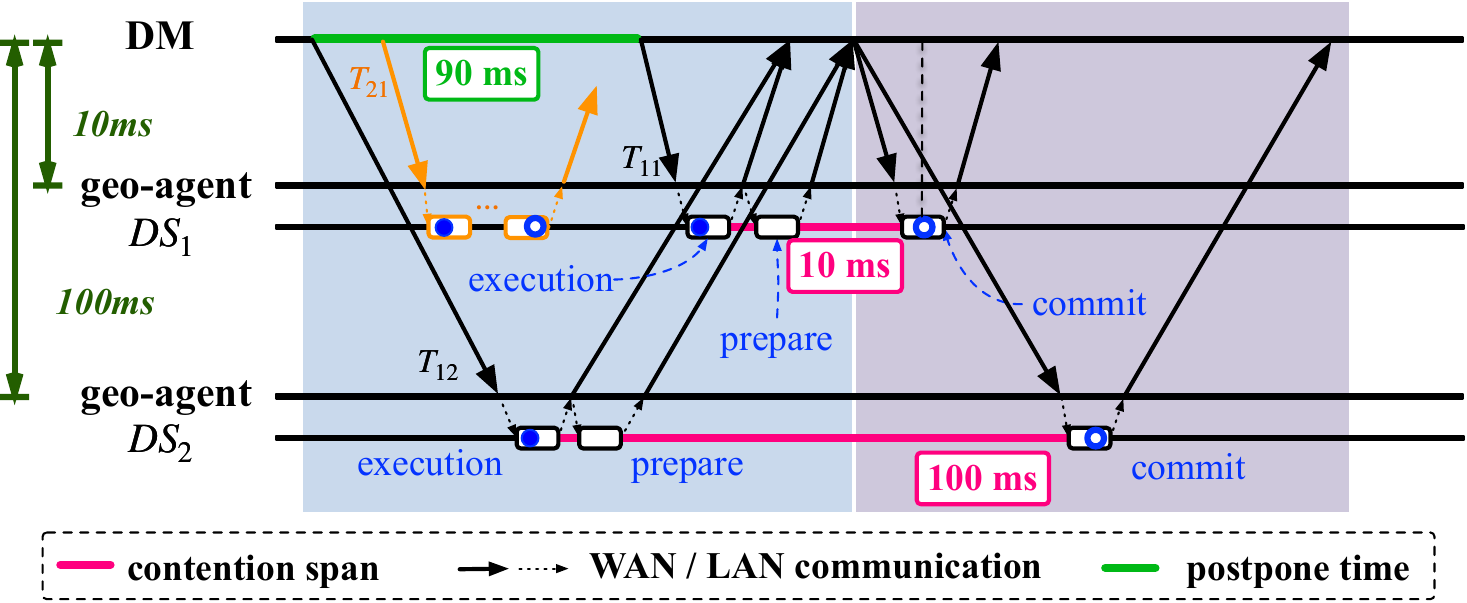}
        \vspace{-6mm}
        \caption{Latency-aware scheduling mechanism}
        \vspace{-2mm}
        \label{design.scheduler}
    \end{subfigure}
    
    \caption{Distributed transaction processing in \dbname}
    \vspace{-7mm}
    \label{design.1}
\end{figure}

\vspace{1mm}
\noindent\textbf{Decentralized prepare mechanism.} 
We propose a decentralized preparation mechanism to offload the coordination cost associated with the prepare phase, i.e., one WAN round trip from the DM to data sources. In \dbname, the prepare phase is initiated by the geo-agent, reducing the cost from one WAN round trip (i.e., from the DM to the data source) to one local-area network (LAN) round trip (i.e., from the geo-agent to the data source). 
The design hint is that once a subtransaction completes its execution phase, it can directly enter the prepare phase without waiting for the \textit{``prepare"} messages from the DM. This approach does not compromise transaction correctness and may incur minimal additional overhead if a client proactively aborts the transaction. However, this overhead is negligible compared to the reduction in WAN round trip time. 
To enable decentralized preparation, we use an annotation code to explicitly mark the last SQL statement in a transaction. Upon completing the execution of the last SQL statement, the geo-agent directly initiates an implicit decentralized prepare phase. For example, if the data source is MySQL, the geo-agent executes \textit{``XA end"} and \textit{``XA prepare"} statements to initiate the prepare phase; for PostgreSQL, it uses the \textit{``Prepared transaction"} statement. Thus, the DM only needs to await the results of this implicit prepare phase before proceeding to the commit phase. Then, the DM flushes the commit/abort log. If all prepared results are successful, the DM notifies involved data sources to commit; otherwise, the DM awaits the abort results from data sources. 
Unlike existing works on reducing round trips~\cite{DBLP:conf/sigmod/YanYZLWSB18, DBLP:conf/sigmod/YangY022, DBLP:journals/pvldb/ZhangLZXLXHYD23}, \dbname does not require modifications to the database kernel, making it suitable for database middleware with heterogeneous data sources.

As illustrated in Figure~\ref{design.agent}, the decentralized prepare mechanism enables the completion of distributed transaction commitments in a single WAN round trip, unlike the traditional two-round trip process.  
The lock contention span for both subtransactions within $T_1$ is reduced from 200 ms (shown in Figure~\ref{back.xa-2pc}) to 100 ms, which corresponds to the longest network round trip time involved in $T_1$. 
\extended{
\begin{algorithm}[t]
    \SetKwInOut{Input}{input}
    \SetKwInOut{Output}{output}
    \caption{Decentralized prepare mechanism}
    \setstretch{0.95}
    \small
    \label{alg.prepare}
    \SetKwFunction{FMain}{AsyncPrepare}
    \SetKwProg{Fn}{Function}{:}{}
    \Fn{\FMain{$T_{ij}$, conn}}{
        result := conn.end($T_{ij}$.Xid) \\
        \If{result \textbf{is} failure}{
            SendMsg($T_{ij}$, \texttt{ROLLBACK\_ONLY}) \\
            AsyncRollback($T_i$, conn) \\
            \Return ERROR\;
        }
        \If{SizeOf($T_{ij}$.peers) \textbf{is} 0}{
            SendMsg($T_{ij}$, \texttt{IDLE}) \textit{// centralized transaction}\\
            \Return SUCCESS;
        }
    
        result := conn.prepare($T_{ij}$.Xid) \\
        
        \If{result \textbf{is} failure}{
            SendMsg($T_{ij}$, \texttt{FAILURE}) \\
            AsyncRollback($T_i$, conn) \\
            \Return ERROR;
        } \Else{
            SendMsg($T_{ij}$, \texttt{PREPARED}) \\
            \Return SUCCESS;
        }
    }

    \SetKwFunction{FMain}{Commit}
    \SetKwProg{Fn}{Function}{:}{}
    \normalem
    \Fn{\FMain{$T_i$}}{
        readyForCommit := true \\

        \textit{/* Block until all prepared results received */} \\
        WaitForPrepareResults() \\
    
        \For{p $\in$ $T_{i}$.participants} {
            \If{p.state \textbf{is not} \texttt{IDLE} \textbf{or} \texttt{PREPARED}}{
                readyForCommit := false \\
            }
        }

        \textit{/* Flush commit/abort log for failure recovery (\S\ref{sec.failure}) */} \\
        FlushLog(readyForCommit) \\
    
        \If{readyForCommit \textbf{is} true}{
            DispatchCommit($T_{i}$.participants) \\
        } \Else {
            WaitForRollback() \\
        }
    }

    \SetKwFunction{FMain}{AsyncRollback}
    \SetKwProg{Fn}{Function}{:}{}
    \normalem
    \Fn{\FMain{$T_{ij}$, conn}}{
        
        \textit{/* Notify other data sources to rollback */} \\    
        \For{p $\in$ $T_{ij}$.peers} {
            NotifyPeerRollback(p.Id, $T_{ij}$.Xid) \\
        }

        conn.rollback($T_{ij}$.Xid) \\
        SendMsg($T_{ij}$.Xid, \texttt{ROLLBACKED})
    }
\end{algorithm}
}

\vspace{1mm}
\noindent\textbf{Early abort mechanism.} 
In conventional DMs, subtransactions remain unaware of the execution status of their peer subtransactions, such as failures due to the lock timeout, until receiving the abort notification from the DM, resulting in resource wastage. 
The challenge is introduced by the inability of heterogeneous data sources to communicate directly with each other. 
Inspired by Guerraoui et.al.~\cite{DBLP:conf/spdp/GuerraouiS95}, \dbname proposes the early abort mechanism to address this issue effectively. In \dbname, the geo-agent maintains connections to other data sources in its connection pool. Once a subtransaction encounters an abort before commitment, the geo-agent proactively notifies other data sources to abort the corresponding peer subtransactions, bypassing the DM and thereby reducing half of the WAN round trip. 
In the previous execution process, when subtransaction $T_{12}$ aborts in $DS_2$, it requires one and a half WAN round trips to abort transaction $T_{11}$ —half a round trip for $DS_2$ to send the abort message to the DM and one round trip for the DM to dispatch the abort command to $DS_1$ and receive the abort result from $DS_1$.
\maintext{
\textcolor{blue}{
An illustration example can be found in the extended version of our paper~\cite{GeoTP}. 
}
}
\extended{
In contrast, as shown in Figure~\ref{desigin.rollback}, \dbname uses the geo-agent in $DS_2$ to proactively notify $DS_1$ directly to abort $T_{11}$. The DM then waits for the abort results from $DS_1$ and $DS_2$, completing the abort of the transaction with only one WAN round trip.
}

\extended{
\vspace{1mm}
\noindent\textbf{Algorithm.} Algorithm~\ref{alg.prepare} details the transaction processing in \dbname. The geo-agent invokes \textit{AsyncPrepare()} to initiate the implicit prepare phase. After the preparation, the geo-agent returns the prepare result to the DM (line 16). When receiving a commit message from the client, the DM invokes \textit{Commit()} to initiate the commit phase. In the commit phase, the DM first waits for all prepare results from data sources and then flushes the commit/abort log (lines 20-26). If all prepare results are successful, the DM invokes \textit{DispatchCommit()} to notify all data sources of the commit message (lines 27-28); otherwise, it invokes \textit{WaitForRollback()} to await the abort results from involved data sources (line 30). Moreover, if a subtransaction encounters an abort before the commitment, its corresponding geo-agent invokes \textit{AsyncRollback()} to perform the early abort. 
}

\subsection{Latency-Aware Scheduling Mechanism\label{design-2}} 

In geo-distributed scenarios, significant differences in network latencies often lead to unnecessary lock contention spans, as we illustrated in the motivation example in Figure~\ref{back.xa-2pc}. To address this issue, we propose a latency-aware scheduling approach to optimize the start time point for each subtransaction. In this part, we assume that there is no data-conflict blocking, allowing subtransactions to acquire locks and complete their execution immediately, for the simplicity of illustration. We will incorporate the transaction execution latency in \S\ref{design-3}. 

\vspace{1mm}
\noindent\textbf{Lock request timing postponing.} 
We first formulate the lock contention span for subtransaction $T_{ij}$ as follows: 
\begin{equation}
\small
    \setlength\abovedisplayskip{4pt}
    \setlength\belowdisplayskip{4pt}
    \label{eq.opt}
    LCS(T_{ij}) = \lastunlocktime - \firstlocktime
\end{equation}
where {\small \firstlocktime} and {\small \lastunlocktime} represent the first lock acquisition time point and the last lock release time point of $T_{ij}$, respectively.



The primary objective of the DM is to minimize each subtransaction's lock contention span defined in Eq.(\ref{eq.opt}). We achieve this by postponing the start time point {\small \transactionstarttime}, which is the time point when the DM dispatches subtransaction $T_{ij}$ to the data source. 
For clarity and without generality, we explain our formulas under the assumptions that (1) the time point when the DM receives the transaction is 0ms, and (2) subtransactions can acquire locks immediately, meaning the first lock acquisition time point {\small \firstlocktime}, can be represented as the time point the target data source receives $T_{ij}$. Formally, {\small \firstlocktime $=$ \transactionstarttime $+ \frac{1}{2}\tau_{ij}$}, where {\small $\tau_{ij}$} denotes the RTT between the DM and the data source where $T_{ij}$ executes.  

Since \dbname eliminates the network round trip in the prepare phase, the last lock release time point {\small \lastunlocktime}, can be represented as the time point when the target data source receives the \textit{``commit"} message for $T_{ij}$. Formally, {\small \lastunlocktime $=$ $\max\limits_{\forall T_{is}\in T_i} \tau_{is} + \frac{1}{2}\tau_{ij}$}, with {\small $\max\limits_{\forall T_{is}\in T_i} \tau_{is}$} representing the highest RTT from the DM to the data sources involved in $T_i$. 
Furthermore, to avoid increasing the overall transaction latency, there is a constraint that the end time point of any subtransaction's execution and prepare phase must not exceed the original end time point of the transaction's entire execution and prepare phase. 
Taking $T_{ij}$ as an example, its end time point of execution and prepare phase can be represented as {\small \transactionstarttime $+ \tau_{ij}$}, the original end time point of $T_i$'s entire execution and prepare phase is {\small $\max\limits_{\forall T_{is}\in T_i} \tau_{is}$}. 
The objective function and constraint (indicated after \textit{s.t.} in Eq.(\ref{eq.opt-1})) for each subtransaction are formally described as follows:
\begin{equation}
\small
    \label{eq.opt-1}
    \setlength\abovedisplayskip{4pt}
    \setlength\belowdisplayskip{4pt}
    \begin{aligned}
        & \mathop{\arg\min}\limits_{\transactionstarttime} LCS(T_{ij}) 
        \Rightarrow \mathop{\arg\min}\limits_{\transactionstarttime} (\max\limits_{\forall T_{is}\in T_i} \tau_{is}  - \transactionstarttime ) \\
        & \text{s.t.}\quad \transactionstarttime + \tau_{ij} \le  \max\limits_{\forall T_{is}\in T_i} \tau_{is} \\
    \end{aligned}
\end{equation}
We can derive the optimal subtransaction start time to minimize each subtransaction's lock contention span in Eq.(\ref{eq.opt-1}) as: 
\begin{equation}
\small
    \setlength\abovedisplayskip{4pt}
    \setlength\belowdisplayskip{4pt}
    \label{eq.solution-1}
    \transactionstarttime = \max_{\forall T_{is}\in T_i} \tau_{is}  - \tau_{ij}
\end{equation}

Notably, for transactions with multiple rounds of interactions, the optimal start time point is calculated for each round.

\begin{algorithm}[t]
    \SetKwInOut{Input}{input}
    \SetKwInOut{Output}{output}
    \caption{Latency-aware scheduling mechanism}
    \label{alg.delay}
    \small
    \SetKwFunction{FMain}{ScheduleTranscation}
    \SetKwProg{Fn}{Function}{:}{}
    \normalem
    \Fn{\FMain{$T_i$}}{
    lat\_max := 0, retry\_cnt := 0 \\
    time\_now := GetSystemClock() \\
   
    \For{$T_{ij} \in T_i$.subtxns}
    {
        node := GetNode($T_{ij}$) \\
        \textit{/* $\tau$ between DM and targeted data source */} \\
        $T_{ij}$.latency := GetNetworkLatency(node)  \\
    }
    \If{adv\_opt \textbf{is} true} {
        \textit{/* Further optimization \S\ref{design-3} */} \\
        \For{$T_{ij} \in T_i$.subtxns}
        {
            p := 0, $\widehat{LEL}(T_{ij})$ := 0 \\
            \For{$r_k$ $\in$ $T_{ij}$.records} {
                    p := UpdatePossibility($r_k$, p) \textit{// Eq.(\ref{eq.abort})} \\ 
                $\widehat{LEL}(T_{ij})$ := $\widehat{LEL}(T_{ij})$ + $w\_lat_{r_k}$
            }
            \If{$p< rand()$}{
                \If{retry\_cnt++ $<$ 10} {
                    \textbf{goto} line 11 \\
                }
                \Return None \textit{/* abort */} \\ 
            }
            \Else{
                $T_{ij}$.latency := $T_{ij}$.latency + $\widehat{LEL}(T_{ij})$ 
                \\
            }
        }
        
    }

    lat\_max := GetMaxSubtransactionLatency($T_i$) \\
    \For{$T_{ij} \in T_i$.sub\_txn}
    {
        \transactionstarttime := time\_now + (lat\_max - $T_{ij}$.latency)
    }
}
\end{algorithm}

\noindent\textbf{Algorithm.} Algorithm~\ref{alg.delay} outlines the key function of the latency-aware scheduling mechanism. \textit{ScheduleTransaction()} is invoked by the geo-scheduler and adjusts the start time point of each subtransaction.
The DM iterates through each subtransaction of the input transaction $T$ and retrieves the network latency between the DM and the target data source (lines 4-7). In this section, we assume the \textit{adv\_opt} as false, with further optimization introduced in \S\ref{design-3} (lines 9-20). 
After that, the DM records the latency of the slowest subtransaction (line 21) and then calculates the optimal start time point for each subtransaction based on Eq.(\ref{eq.solution-1}) (lines 22-23). 
Recall the example transaction in Figure~\ref{design.agent}, the distributed transaction $T_2$ arrives at 5 ms in the DM and needs to access the same record $r$ on $DS_1$ with transaction $T_1$. Without postponing the start time point of $T_{11}$, $T_{21}$ needs to wait until 105 ms, when $T_{11}$ releases its locks. In \dbname, as shown in Figure~\ref{design.scheduler}, our geo-scheduler postpones the start time point of subtransactions $T_{11}$ by 90 ms, $T_2$ can acquire the locks ahead of $T_{11}$ and release locks before $T_{11}$ arrives. 
Consequently, the lock contention span for subtransactions $T_{11}$ and $T_{12}$ and $T_{21}$ are reduced to 10 ms, 100 ms and 10ms, respectively. 
This postponing mechanism enhances transaction concurrency, leading to an overall improvement in system performance. 



\subsection{High-Contention Workload Optimizations}
\label{design-3}

The previous discussion assumed that the transactions are under low-contention workloads. However, in high-contention workloads, the lock contention span is influenced not only by the longest RTT but also by the time required for subtransactions to acquire locks. In high-contention workloads, subtransactions often cannot immediately acquire locks. Additionally, frequent transaction waits or rollbacks due to the contention waste system resources and significantly undermine the effectiveness of predicting the time required for acquiring locks.
To address the aforementioned challenges, we propose a heuristic \textit{local execution latency forecasting} mechanism to improve latency-aware scheduling using real-time statistical information. Additionally, we introduce a \textit{late transaction scheduling} mechanism to manage access to hot records. 




\vspace{1mm}
\noindent\textbf{Hotspot statistics collecting.} Since the \textit{local execution latency forecasting} and \textit{late transaction scheduling} rely on real-time hotspot statistics, we first introduce how \dbname collects this information.  
The geo-scheduler uses the hotspot footprint to maintain statistics for hot records of data sources, including four fields: (1) {\small $w\_lat_{r}$}, the weighted average latency of subtransactions completing operations on the record {\small $r$}; (2) {\small $t\_cnt_{r}$}, the total number of transactions that have accessed the record {\small $r$}; (3) {\small $c\_cnt_r$}, the number of committed transactions that have accessed the record {\small $r$}; (4) {\small $a\_cnt_r$}, the number of transactions currently accessing the record {\small $r$}. 
We update these fields after the completion of each subtransaction $T_{ij}$ within transaction $T_i$. 
Specifically, to update the {\small $w\_lat_{r}$} for the record {\small $r$} that $T_{ij}$ has accessed, the DM uses a weighted average approach as formulated in Eq.(\ref{eq.statistics}). 
Since the latency of $T_{ij}$ accessing a specific record $r$ cannot be directly collected (due to data record granularity), we calculate it using a weight {\small $w_r=\frac{w\_lat_{r}}{\sum_{\forall r_k \in T_{ij}.records}{w\_lat_{r_k}}}$}, the proportion of {\small $w\_lat_{r}$} relative to the sum of access latency of all records accessed by $T_{ij}$. 
The latency of $T_{ij}$ accessing the record $r$ is then estimated by {\small $LEL(T_{ij}) \cdot w_r$,} with {\small $LEL(T_{ij})$} denoting the local execution latency of $T_{ij}$. Then we use this latency to update {\small $w\_lat_r$}, with the weighted update coefficient {\small $\alpha$}. 
\begin{equation}
\small
    \label{eq.statistics}
    w\_lat_{r} = \alpha \cdot w\_lat_{r} + (1 - \alpha) \cdot LEL(T_{ij}) \cdot w_r \\
\end{equation}

To enhance efficiency, we organize these hot records using an AVLTree in memory, which ensures that both point and range access has a time complexity of $O(\log n)$. Additionally, we implement an LRU list to evict cold data, allowing \dbname to dynamically update hot records during operation. This approach not only reduces the memory overhead but also minimizes the CPU overhead for latency estimation. 

\vspace{1mm}
\noindent\textbf{Local execution latency forecasting.} As formulated in Eq.(\ref{eq.prediction}), based on the collected statistical information on hot records, we estimate the local execution latency of subtransaction $T_{ij}$ by accumulating the value of $w\_lat_{r}$ of each hot record that $T_{ij}$ need to access. 
To distinguish from the actual local execution latency of $T_{ij}$, we use {\small$\widehat{LEL}(T_{ij})$} to represent the forecasted local execution latency. 
{
\begin{equation}
\small
    \label{eq.prediction}
    \widehat{LEL}(T_{ij}) = \sum\nolimits_{\forall r_k \in T_{ij}.records} {w\_lat_{r_k}} \\
\end{equation}
}

Then we incorporate the forcasted local execution latency into Eq.(\ref{eq.opt}). 
The {\small \firstlocktime} is updated by adding {\small $Req(r_{1st})$}, the time span from the lock request to the lock acquisition on $T_{ij}$'s first accessing record $r_{1st}$. The {\small \lastunlocktime} is updated by adding the forecasted local execution latency to the execution and prepare phase of subtransactions.
The updated {\small \firstlocktime} and {\small \lastunlocktime} are formulated in Eq.(\ref{eq.lutime}). 
\begin{equation}
\label{eq.lutime}
\setlength\abovedisplayskip{4pt}
\setlength\belowdisplayskip{4pt}
\small
    \begin{aligned}
        \firstlocktime &= \transactionstarttime + \tfrac{1}{2}\tau_{ij} + Req(r_{1st}) \\
        \lastunlocktime &= \max\limits_{\forall T_{is}\in T_i} 
        (\tau_{is} + \widehat{LEL}(T_{is}))  + \tfrac{1}{2}\tau_{ij} \\
    \end{aligned}
\end{equation}
Finally, we generate the new objective function and constraint for the lock contention span using Eq.(\ref{eq.lutime}).
\begin{equation}
    \label{eq.objective-constraint}
    \setlength\abovedisplayskip{4pt}
    \setlength\belowdisplayskip{4pt}
    \small
    \begin{aligned}
        & \mathop{\arg\min}\limits_{\transactionstarttime} LCS(T_{ij}) \\
        \Rightarrow & \mathop{\arg\min}\limits_{\transactionstarttime} [\max\limits_{\forall T_{is}\in T_i} (\tau_{is} + \widehat{LEL}(T_{is})) - (\transactionstarttime + Req(r_{1st}))] \\
        & \text{s.t.}\quad \transactionstarttime + \tau_{ij} + \widehat{LEL}(T_{ij}) \le \max\limits_{\forall T_{is}\in T_i} (\tau_{is} + \widehat{LEL}(T_{is})) \\
    \end{aligned}
\end{equation}
Since {\small $Req(r_{1st})$} is contained in {\small $\widehat{LEL}(T_{ij})$}, it can be considered as a constant without affecting the optimal solution. 
Therefore, the optimal start time point can be formulated as follows: 
\begin{equation}
\small
    \label{eq.solution-2}
    \setlength\abovedisplayskip{3pt}
    \setlength\belowdisplayskip{3pt}
    \transactionstarttime = \max_{\forall T_{is}\in T_i} (\tau_{is} + \widehat{LEL}(T_{is})) - (\tau_{ij} + \widehat{LEL}(T_{ij}))
\end{equation}

Moreover, discrepancies between predicted and actual latency do not always degrade performance.
According to Eq.(\ref{eq.solution-2}), if the predicted latency $\widehat{LEL}(T_{ij})$ is lower than the actual latency $LEL(T_{ij})$, Eq.(\ref{eq.objective-constraint}) may not achieve the minimal value, but still perform better than execution without latency-aware scheduling. However, if $\widehat{LEL}(T_{ij})$ exceeds $LEL(T_{ij})$, performance may suffer if the delayed subtransaction becomes the new bottleneck. In cases of inaccurate runtime predictions, we can scale down the predicted latency before incorporating it into calculations to mitigate any negative impact.

\vspace{1mm}
\noindent\textbf{Late transaction scheduling.} To restrict the number of concurrent transactions on hot records and improve prediction accuracy, the DM first predicts the abort rate of transactions before distributing them to data sources, blocking those with high abort rates. 
Specifically, a transaction will be aborted if it cannot acquire locks on any of the records due to lock timeout. Therefore, the transaction's abort rate, denoted as {\small $Pr(T_i)$}, is equivalent to 1 minus the probability of the transaction successfully acquiring locks on all required records. 
This prediction is conducted with the hotspot footprint. We observe that if a transaction is blocked by another transaction with a high abort rate, other transactions waiting for the blocked transactions are also likely to be aborted. Therefore, we predict the probability that a transaction can successfully acquire the lock on a record by calculating the probability that all preceding transactions in the waiting queue can successfully acquire locks. The number of transactions in the waiting queue can be represented by  {\small $a\_cnt_{r_k}-1$} and each transaction has a {\small $\frac{c\_cnt_{r_k}}{t\_cnt_{r_k}}$} probability of successfully acquiring the lock on $r_k$ without being blocked. 
Given that, we formalize the abort rate for transaction $T_i$ in Eq.(\ref{eq.abort}):
\begin{equation}
\small
    \label{eq.abort}
    \setlength\abovedisplayskip{3pt}
    \setlength\belowdisplayskip{3pt}
    Pr(T_i) = 1 - \prod_{\forall r_k \in T_i.records}{(\frac{c\_cnt_{r_k}}{t\_cnt_{r_k}})^{\max{\{a\_cnt_{r_k} - 1, 0\}}}}
\end{equation}

\vspace{1mm}
\noindent\textbf{Algorithm.} We integrate the late transaction scheduling and local execution latency forecasting mechanisms for further optimization, as illustrated in Algorithm \ref{alg.delay}. 
When the adv\_opt is true, we consider the local execution latency via \textit{PredictLatency()}.
Specifically, the DM traverses the keys accessed by the subtransactions and predicts the abort rate and local execution latency based on Eq.(\ref{eq.prediction}) and Eq.(\ref{eq.abort}) (lines 13, 20). Transactions with a high abort rate are blocked (line 17). Otherwise, the DM calculates the optimal start time point for each subtransaction based on the network latency and predicted local execution latency (line 23). Additionally, the transactions that have been blocked multiple times are aborted (line 18). 

\subsection{Discussion} \label{design.discussion}
The high-contention optimization estimates the local execution latency. 
When a user specifies a predicate on the primary key or a key from a secondary index, we can identify hot records cached in the memory that match the predicate, enabling us to estimate the local execution latency of subtransactions. 
In certain cases, such as when there is no index on the predicate key, inferring hot records from the statements becomes challenging and inefficient, which may limit the effectiveness of this technique. However, other optimizations continue to enhance performance. 
Additionally, dependencies between operations from different subtransactions may necessitate multiple rounds of interactions. However, within each round, the geo-scheduler can optimize the scheduling of query execution. 
\maintext{
\textcolor{blue}{
Due to space limitations, we discuss the integrity constraint problems in our extended versions~\cite{GeoTP}.
}
}
\extended{
Furthermore, integrity constraints within the database may exist. While deferring the execution of a subtransaction that violates these constraints may not always improve performance, it can delay the client's perception of the error, potentially hindering feedback regarding transaction execution failure. From the perspective of overall system performance, the benefits of latency awareness outweigh these drawbacks.
For scenarios where integrity errors occur frequently, a heuristic algorithm could be developed to defer subtransactions involving those records with a small probability. This intriguing problem, however, lies beyond the scope of this paper and is left as a topic for future research.
}

\section{Correctness and Recovery}
In this section, we first describe the failure recovery mechanism of \dbname. Then, we provide proofs of the atomicity and isolation correctness in \dbname.



\subsection{Failure Recovery \label{sec.failure}}


The failure recovery process of \dbname includes three key aspects: (1) identifying which transactions require recovery, (2) determining where to collect the necessary information for recovery, and (3) deciding how to recover these transactions. We discuss the recovery process for both the DM failures and data source failures. 
Note that our recovery process relies on the following common settings in popular databases~\cite{mysql, postgresql}: \blackding{1} if the DM fails and disconnects, the data sources abort all subtransactions not completed the prepare phase; 
and \blackding{2} if a data source fails, it automatically aborts subtransactions that have not completed the prepare phase after it restarts.

\maintext{
When the DM encounters a failure, \dbname would abort uncommitted transactions that have not entered the commit phase and then complete uncommitted transactions that have once it restarts. On the other hand, when a data source, such as $DS_i$, fails along with its geo-agent. In this case, the DM is responsible for recovering all uncommitted distributed transactions that accessed this failed data source. The DM aborts the distributed transaction if its subtransaction in $DS_i$ has not completed the prepare phase. Otherwise, the DM continues to execute the distributed transaction after the data source is reconnected. 
{\color{blue}
The detailed recovery process can be found in our extended versions \cite{GeoTP}.
}
}

\extended{
\vspace{1mm}
\noindent\textbf{Recover from database middleware failure.} 
When the DM encounters a failure, \dbname recovers all transactions that were not committed before the failure. 
Particularly, the DM needs to abort uncommitted transactions that have not entered the commit phase and complete uncommitted transactions that have entered the commit phase. 
Since the DM is stateless, it retrieves the transaction information from the underlying data sources after it restarts. 
First, the DM reconnects to data sources and collects all prepared but uncommitted subtransactions to identify all uncommitted distributed transactions. Second, the DM uses the persisted commit/abort log entry to check if transactions entered the commit phase. If a log entry exists, indicating that this transaction 
has completed the \textit{FlushLog()} in Algorithm~\ref{alg.prepare} 
and has entered the commit phase, the DM instructs the data source to commit/abort the corresponding subtransactions based on the log; If not, the DM aborts the transaction. 

\vspace{1mm}
\noindent\textbf{Recover from data source failure.}
When a data source, such as $DS_i$, fails, the corresponding geo-agent also fails. In this case, the DM is responsible for recovering all uncommitted distributed transactions that accessed this failed data source. The DM will abort the distributed transaction if its subtransaction in $DS_i$ has not completed the prepare phase. Otherwise, the DM continues to execute the distributed transaction after the data source is reconnected. 
Specifically, the DM first traverses each distributed transaction involving $DS_i$ to retrieve the transaction state. If the transaction has reached the commit phase, indicating that it has completed the prepare phase in $DS_i$, the DM instructs $DS_i$ to either commit or abort the corresponding subtransaction based on the commit/abort log of this transaction. If the subtransaction is not in the commit phase, the DM communicates with the geo-agent to verify whether the subtransaction in $DS_i$ has been successfully prepared. If it has, the database middleware processes the subtransaction as per the normal procedure. Otherwise, the modification is lost in $DS_i$ and the DM will notify all data sources involved in the distributed transaction to abort the corresponding subtransactions.
}

\subsection{Atomicity Correctness}
To ensure atomicity, we must guarantee that the final state of the transaction is either committed or aborted, with all subtransactions achieving the same status.  
\dbname guarantees the atomicity correctness following two steps. First, \dbname initiates the decentralized prepare phase after the execution phase. A transaction can be committed only if all subtransactions complete the prepare phase and vote \textit{Yes}; otherwise, the transaction will be aborted. Second, once the final status of the transaction is determined, \dbname flushes the commit/abort log into the disk, ensuring that the decision cannot be reversed. 
In the case of failures, we ensure that the DM and the underlying data sources eventually reach a unique and consistent decision after failure recovery (\S\ref{sec.failure}).

\extended{
We next prove atomicity of \dbname, referring to~\cite{DBLP:books/aw/BernsteinHG87, DBLP:journals/pvldb/Kraft0ZBSYZ23} for five properties (AC1-5) that \textit{atomic commit protocols} need to fulfill.

\textbf{AC1: All processes that reach a decision reach the same one.} In the commit phase, the DM determines the final state of the transaction according to the prepared results and dispatches the decision to each data source. The DM can commit only if all data sources are ready to commit. 

\textbf{AC2: A process cannot reverse its decision after it has reached one.} Data sources can not reverse the decision once it commits/aborts the transaction. For DM, it flushes the commit/abort log after it determines the final state. \dbname would reuse this decision (if necessary) rather than reverse it.

\textbf{AC3\&4: The Commit decision can only be reached if all processes voted Yes. If there are no failures and all processes voted Yes, then the decision will be to Commit.} \dbname adheres to the decision logic of the 2PC (details in \S\ref{design-1}), simply reducing the network overhead under WAN. Therefore, it meets this property.

\textbf{AC5: Consider any execution containing only failures that the algorithm is designed to tolerate (i.e., crash failures). At any point in this execution, if all existing failures are repaired and no new failures occur for sufficiently long, then all processes will eventually reach a decision.}
In the previous description, \dbname ensures that the data sources can accurately execute the decision made by the middleware. \S\ref{sec.failure} will discuss how \dbname handles failures at different times in the middleware and data sources, thereby ensuring that the data source applies the unique decision made by the DM in the event of a failure.
}

\subsection{Isolation Correctness}
\dbname is an effective distributed transaction processing approach that can be applied to multiple middleware systems. \dbname postpones the execution of subtransactions but does not modify the concurrency control algorithm, thereby maintaining the isolation properties of the original middleware system. Take the 2PL algorithm as an example, latency-aware scheduling postpones the acquisition of locks but still requires the acquisition of a lock before its operation, as enforced by the concurrency control mechanism of each data source. The commit protocol of \dbname ensures that a transaction can release locks only after commitment, thereby preserving serializability. Thus, \dbname does not compromise the isolation guarantees the original database middleware provided. 

\section{IMPLEMENTATION\label{sec.implementation}} 
We implement \dbname on the codebase of   
Apache Shardingsphere-v5.0 \cite{DBLP:conf/icde/LiZPLWSWCGG22}, a popular open-source DM, involves about 5k lines of Java code modifications and is available at~\cite{GeoTP}. 
\extended{
Shardingsphere currently integrates 6 popular databases and deploys a comprehensive SQL engine. 
}
\dbname does not require any modifications to data sources. As a result, databases supported by Shardingsphere (possibly other DMs) can leverage the capabilities of \dbname. 

\noindent\textbf{Database middleware layer.} First, we enhance the \textit{sqlParse()} to recognize the annotation code provided by the client and convey this information via \textit{QueryContext}. Then, we enhance the statement handler to support latency-aware scheduling facilitated by the geo-scheduler. The statistical data required for scheduling is derived from two sources: (1) a dedicated thread that continuously monitors the network latency between the DM and data sources, utilizing the \textit{ping} command at 10ms intervals, and (2) the hotspot footprint is recorded by {\textit{LockMetaTable}} and updated by \textit{MultiStatementsHandler.feedback()}. Lastly, we implement the decentralized prepare mechanism in \textit{XAShardingSphereTransactionManager.commit()} to eliminate one WAN round trip for distributed transactions. 

\noindent\textbf{Data source layer.} We have implemented the geo-agent in the data source layer, which 
incorporates an enhanced connection pool and a local transaction manager. 
The connection pool is designed to interface with other geo-agents, thereby supporting functionalities such as {\textit{SendRollbackMsg()}} and {\textit{SendAsyncMsg()}}.
The transaction information is systematically organized within the local transaction manager using a \textit{ConcurrentHashMap}. Before the return of {\textit{CommandExecutorTask.run()}}. We invoke the {\textit{AsyncPrepare()}} in an asynchronous thread upon identifying that the `last' flag is true and then start the prepare phase of the corresponding subtransaction.
\extended{
In particular, the geo-agent autonomously performs \textit{FastRollback()} in failure cases without relying on instructions from the DM.
}

\section{Performance Evaluation\label{sec.evaluation}}

\subsection{Experiment Setup} 
\label{sec.evaluation.setup}

We conduct experiments on an in-house cluster of up to 6 separate machines,  each equipped with 16 vCPUs and 32 GB of DRAM, running CentOS 7.4. 


\subsubsection{Baselines}
In our experiments, we compare \dbname with state-of-the-art DMs and a distributed database.

\vspace{1mm}
\noindent\textbf{Database middlewares. }
\whiteding{1} Shardingsphere (abbreviated as SSP) is a state-of-the-art DM that supports geo-distributed transaction processing over relational database systems via XA protocol.
\whiteding{2} SSP (local), a mode provided by SSP without atomicity guarantees,
which we employ to show the peak performance of SSP. Succinctly, it employs a decentralized commit protocol but allows transactions to be committed when data sources return different votes.
\whiteding{3} ScalarDB~\cite{DBLP:journals/pvldb/YamadaSIN23}, another state-of-the-art DM, supports geo-distributed transaction processing without specific requirements for underlying data sources. 
\whiteding{4} ScalarDB+, a variant of ScalarDB, we implemented by integrating the latency-aware transaction scheduling mechanism and the heuristic optimization. We use ScalarDB+ to study the scalability of our proposed approach.

\vspace{1mm}
\noindent\textbf{Distributed databases. } 
\whiteding{5} YugaByteDB, an advanced distributed database that supports intelligent data partitioning and geo-distributed transactions. 

\vspace{1mm}
\noindent\textbf{Transaction scheduling techniques. } 
\whiteding{6} QURO, a transaction preprocessing technique, reorders write operations as late as possible within the transaction. 
\whiteding{7} Chiller, a distributed transaction protocol, eliminates the prepare phase and schedules the cross-region subtransactions after the intra-region ones are complete. 
For a fair comparison, we implemented both techniques on the same platform as \dbname.

\subsubsection{Benchmarks} 
We adopt the following two benchmarks.

\noindent\textbf{YCSB}~\cite{DBLP:conf/cloud/CooperSTRS10, DBLP:conf/icde/DeyFNR14} generates synthetic workloads that simulate large-scale Internet applications. 
We use the YCSB transactional variant adopted in related works~\cite{DBLP:journals/pvldb/ZhangLZXLXHYD23, DBLP:journals/tkde/ZhaoZZLLZPD23}, where each transaction has 5 operations by default, each with a 50\% probability of being a read or write. 
We run the workloads on a table partitioned with 1 million records per data node. 
Each record consumes 1KB, culminating in a total of 4GB of data hosted by the table.
We control the distribution of accessed records using the parameter \textit{skew\_factor}, where a higher \textit{skew\_factor} results in greater contention.
We set the skew factor to 0.3, 0.9, and 1.5 for low, medium, and high-contention workloads.


\noindent\textbf{TPC-C}~\cite{TPCC} is a popular OLTP benchmark modeling a warehouse order processing application.
The workloads consist of 9 relations, with each warehouse being 100 MB in size. 
By default, each data node hosts 16 warehouses.
In our experiments, we use the standard TPC-C with 5 types of transactions by default. Following previous works~\cite{DBLP:journals/pvldb/YuBPDS14, DBLP:journals/pvldb/HardingAPS17}, we exclude `think time' and user data errors that cause 1\% of NewOrder transactions to abort.




\subsubsection{Default Configuration}
We use one machine to serve as the client, generating transaction requests by Benchbase \cite{DBLP:journals/pvldb/DifallahPCC13}.
By default, we run 64 client terminals. 
For YCSB, we use the medium contention by default.
The default ratio of distributed transactions is set to 0.2. 
For the remaining 5 machines, we deploy database middleware on 1 machine, while the other 4 machines are data nodes.
We emulate the default geo-distributed network environment via \textit{tc} command~\cite{TC}. 
The client, \dbname, and a data node are located in Beijing, with the other data nodes in Shanghai, Singapore, and London. 
Based on the network evaluation conducted in the cloud, the average latency between the corresponding data nodes and \dbname are 0ms, 27ms, 73ms, and 251ms, respectively. 
Unless otherwise specified, we use the default settings to conduct our experiments.
The remaining 4 nodes host MySQL v8.0.22 and PostgreSQL v15.2 as data nodes. 
Except for \S\ref{sec.eva.heterogeneous}, where we specify otherwise, all data nodes run MySQL v8.0.22 as data sources.
By default, we set the isolation level to serializable 
\extended{
for DMs and the data sources (MySQL/PostgreSQL) 
}
and configure the buffer pool size to 24GB and the lock-wait timeout to 5s. The rewriter in middleware will replace \textit{SELECT} with \textit{SELECT...FOR SHARE} to add an explicitly shared lock for PostgreSQL's read operations. 
\extended{
All other settings of DMs and data sources are kept to their default values unless specifically stated.
}


\subsection{Overall Performance of \dbname \label{sec:evaluation_scalability}}

\begin{figure}[t]
    \centering
    \begin{minipage}{0.8\linewidth}
        \centering
        \includegraphics[width=\linewidth]{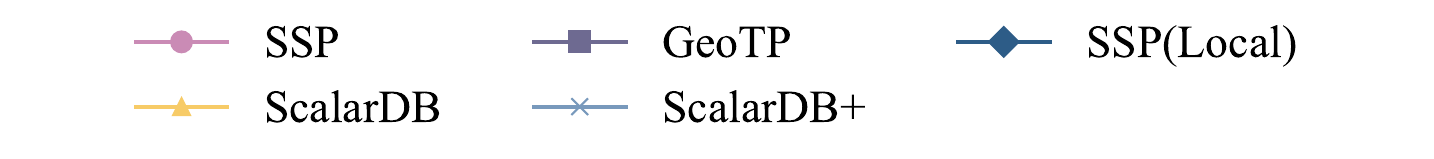}
        \vspace{-5mm}
    \end{minipage}

    \begin{subfigure}{0.48\linewidth}
        \includegraphics[width=\linewidth]{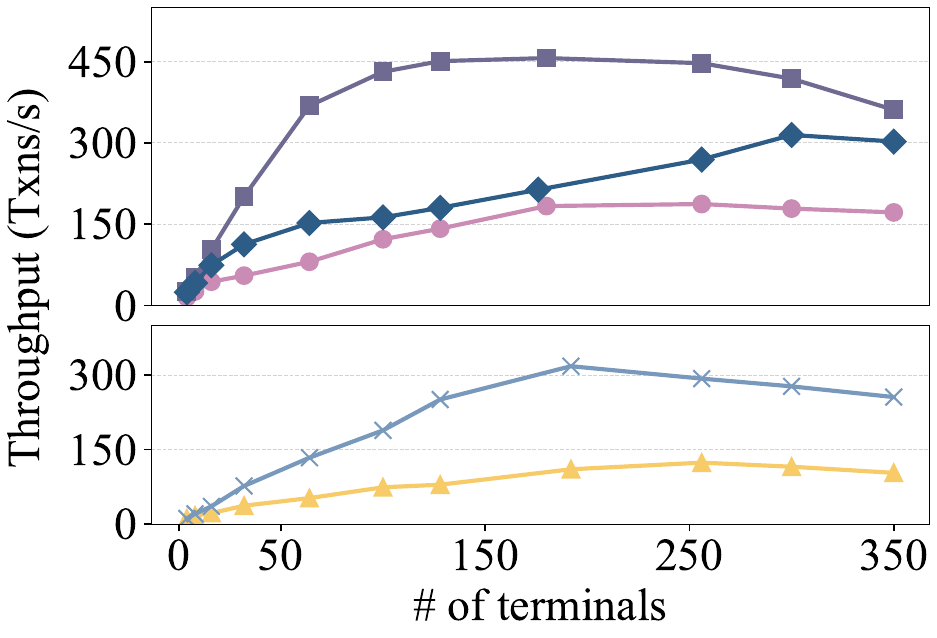}
        \vspace{-6mm}
        \caption{Scalability - YCSB}
        \label{Fig.ycsb.sca.per}
    \end{subfigure}
    \begin{subfigure}{0.48\linewidth}
        \includegraphics[width=\linewidth]{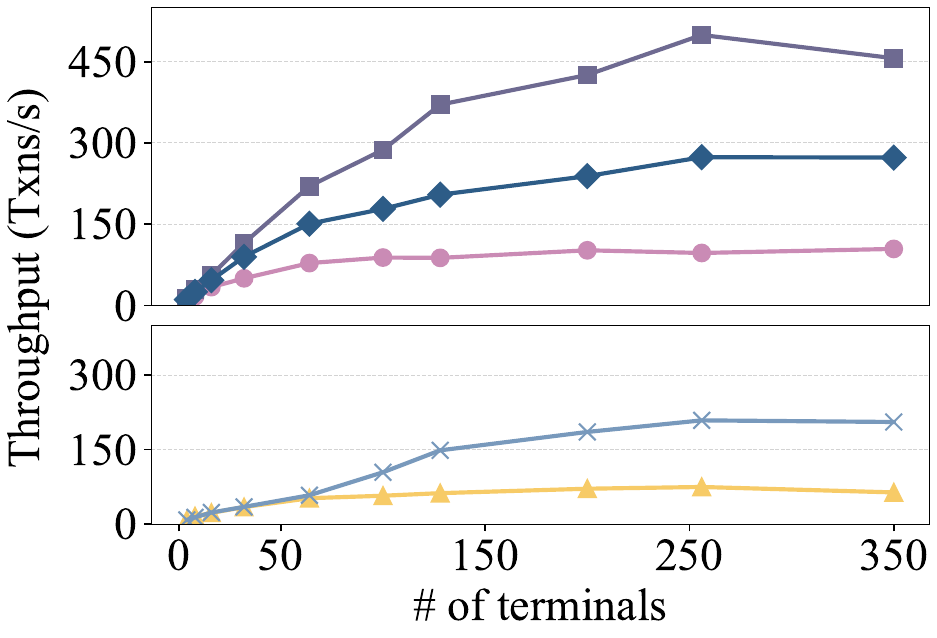}
        \vspace{-6mm}
        \caption{Scalability - TPC-C}
        \label{Fig.tpcc.sca.per}
    \end{subfigure}
    \vspace{-2mm}
    \caption{Overall performance comparison 
    }
    \label{Fig.experiment.sca}
    \vspace{-5mm}
\end{figure}

We first compare \dbname with state-of-the-art DMs using YCSB and TPC-C benchmarks with varying numbers of client terminals.
As shown in Figure~\ref{Fig.experiment.sca}, \dbname outperforms SSP(Local), SSP, and ScalarDB by up to 2.65x, 5.14x, and 7.15x, respectively. 
This throughput improvement is attributed to the latency-aware and late transaction scheduling of \dbname, which effectively reduces lock contention span and enhances concurrency. 
ScalarDB, on the other hand, does not rely on the transactional capabilities of underlying data sources but solely on DM nodes for concurrency control, which limits its scalability and performance.
Moreover, we can observe that ScalarDB+ achieves up to 3.16x and 3.22x throughput gain over ScalarDB under YCSB and TPC-C, respectively.
This demonstrates the general applicability of the proposed techniques in \dbname.  
As the number of terminals increases, all approaches experience a decline in performance. This decline is attributed to system resource competition and lock contention within the databases. 

\begin{figure}[t]
    \centering
    \begin{subfigure}{0.3\linewidth}
        \includegraphics[width=\linewidth]{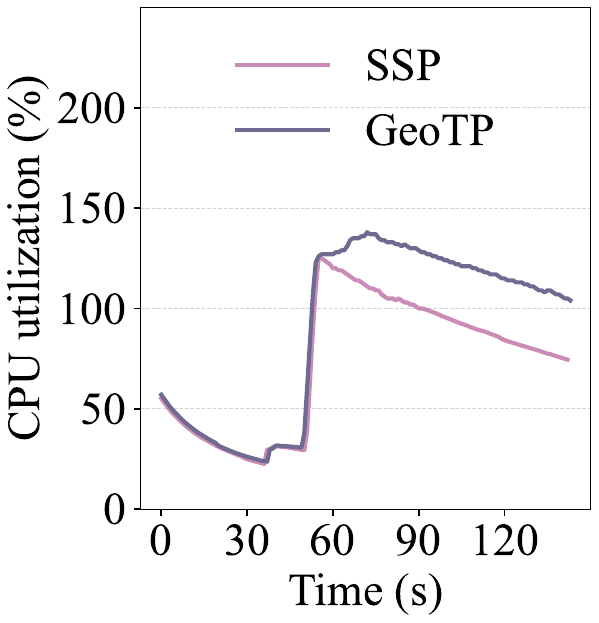}
        \vspace{-6mm}
        \caption{CPU}
        \label{Fig.stat.cpu}
    \end{subfigure}
    \begin{subfigure}{0.3\linewidth}
        \includegraphics[width=\linewidth]{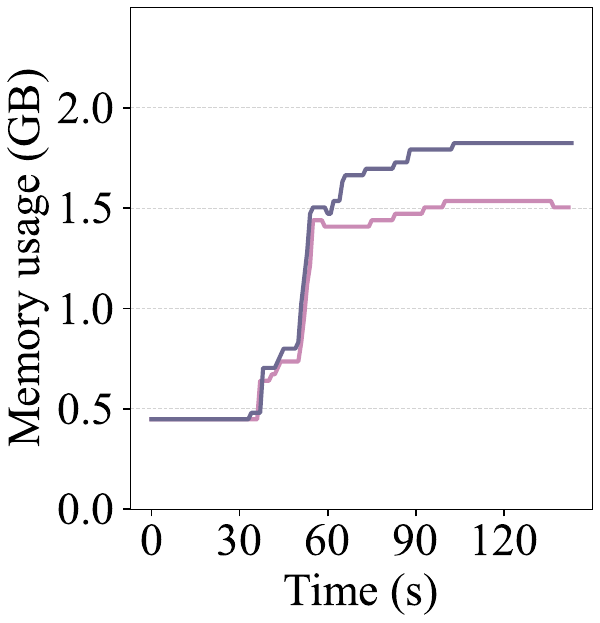}
        \vspace{-6mm}
        \caption{Memory}
        \label{Fig.stat.mem}
    \end{subfigure}
    \begin{subfigure}{0.3\linewidth}
        \includegraphics[width=\linewidth]{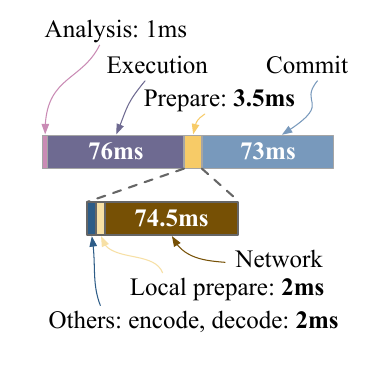}
        \vspace{-6mm}
        \caption{Breakdown}
        \label{Fig.stat.breakdown}
    \end{subfigure}
    \vspace{-1mm}
    \caption{Resource Utilization over YCSB}
    \label{Fig.experiment.overhead}
    \vspace{-6mm}
\end{figure}

We analyze the overhead of \dbname, with results shown in Figure \ref{Fig.experiment.overhead}. Figure \ref{Fig.stat.cpu} presents CPU utilization, where \dbname achieves around 30\% higher CPU efficiency due to network latency detection and latency-aware scheduling. Figure \ref{Fig.stat.mem} shows memory usage, with \dbname requiring approximately 300MB more than SSP as it maintains metadata for hot records in memory. Lastly, Figure \ref{Fig.stat.breakdown} provides a breakdown of module costs in a single transaction lifecycle. Here, analysis overhead remains under 1 ms, with a 3.5 ms wait to enter the commit phase, facilitated by the decentralized prepare mechanism. Compared to the entire transaction latency, the overhead incurred by \dbname is negligible, while \dbname achieves 66.6\% lower average latency compared to SSP.

\subsection{Impact of Distributed Transactions}\label{sec:evaluation_workloads}
We now compare \dbname against other baselines by varying the percentage of distributed transactions. 

\begin{figure}[t]
    \centering
    \begin{minipage}{0.4\textwidth}
        \centering
        \includegraphics[width=\linewidth]{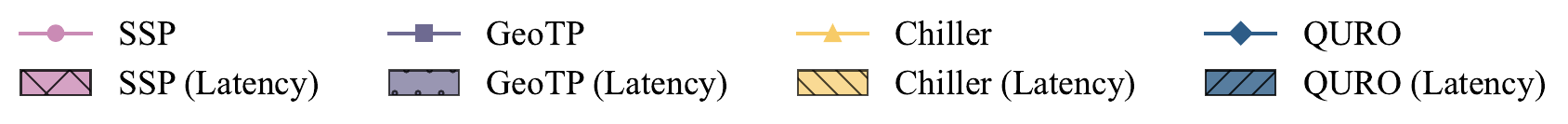}
        \vspace{-5mm}
    \end{minipage}

    \begin{minipage}{0.46\textwidth}
        \centering
        \includegraphics[width=\linewidth]{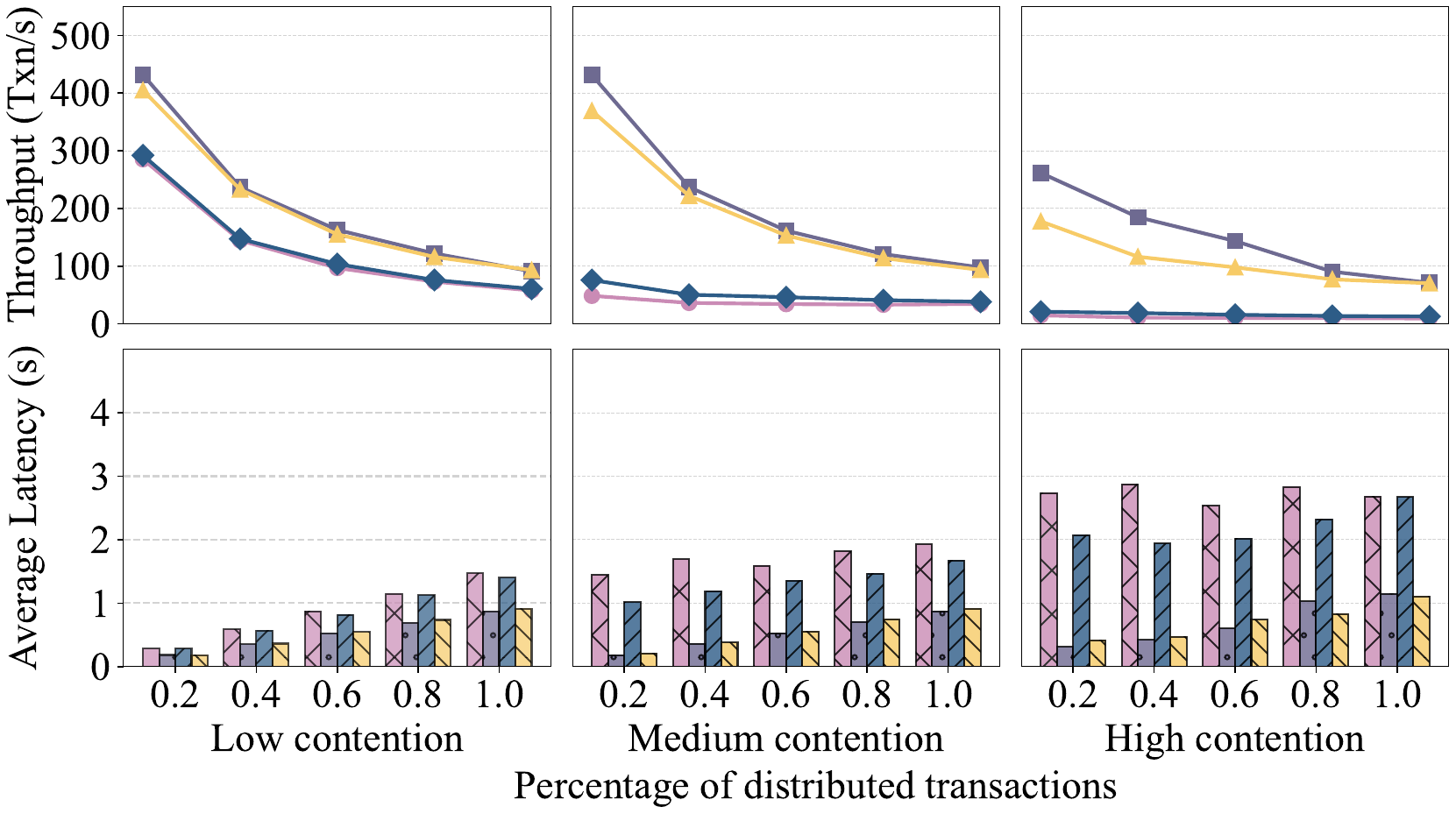}
        \vspace{-4mm}
    \end{minipage}
    \vspace{-2mm}
    \caption{Impact of distributed transactions over YCSB}
    \label{Fig.experiment.ycsb.ds}
    \vspace{-5mm}
\end{figure}

\vspace{0.5ex} \noindent \textbf{YCSB}: 
We control the ratio of distributed transactions by generating keys assigned to different data nodes. We evaluate \dbname under three levels of contention as used in Figure~\ref{Fig.experiment.ycsb.ds}. \dbname outperforms in all three scenarios. 
As the proportion of distributed transactions increases, although the performance of both systems declines, the advantages of \dbname are still pronounced. 
\dbname outperforms Chiller up to 1.6x and other baselines up to 8.9x.
The evaluation results show that while QURO performs better than SSP, it still falls short compared to other methods because it doesn't consider network latency, making it unsuitable for geo-distributed scenarios. Chiller merges the prepare phase with execution and executing inner-region subtransactions after outer-region ones. This approach reduces the lock contention span in low and medium workloads, allowing Chiller to achieve comparable performance. However, \dbname includes specific optimizations for high-contention workloads, further enhancing its performance. It is aligned with our findings discussed in \S~\ref{sec:evaluation_optimization}.


\begin{figure}[t]
    \centering
    \includegraphics[width=\linewidth]{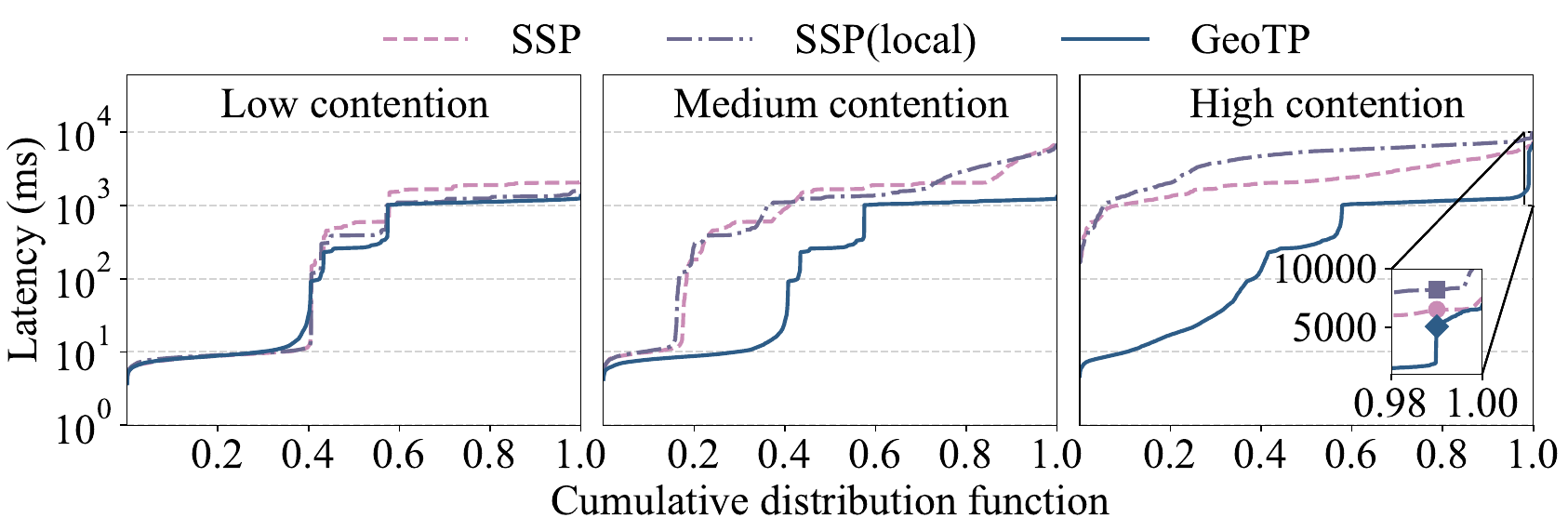}
    \vspace{-6mm}
    \caption{Analysis of latency CDF over YCSB}
    \label{Fig.experiment.ycsb.ds.cdf}
    \vspace{-5mm}
\end{figure}

We further analyze the latency distribution of transactions (with 60\% of distributed transactions) in three scenarios using Cumulative Distribution Function (CDF) plots, as depicted in Figure~\ref{Fig.experiment.ycsb.ds.cdf}. 
\dbname consistently reduces the latency of distributed transactions across all three scenarios. We use \textbf{turning point} to describe the point where transaction latency experiences a significant increase.
In the low-contention workload, with a turning point at 0.4, the latency of 40\% centralized transactions remains unaffected by distributed transactions. 
As contention increases in the medium-contention workload, the turning point of SSP and SSP(local) occurs at about 0.2, which suggests that around 20\% of the execution latency in centralized transactions experiences an increase due to the distributed transaction. 
In contrast, the turning point of \dbname remains around 0.4, with a 99th-percentile (p99) latency lower than the baselines up to 35.9\%. 
This improvement is due to our latency-aware scheduling, which mitigates lock contention and lessens the impact of distributed transactions on centralized transactions. 
In high-contention workloads, SSP and SSP(local) show a turning point near 0, reflecting severe latency spikes in centralized transactions. However, \dbname exhibits no distinct turning point; instead, its latency rises gradually and remains considerably lower overall. Although \dbname maintains lower p99 latency, its p99.9 advantage diminishes due to the optimization in \S~\ref{design-3}, which can increase latency by introducing blocks and aborts.


\begin{figure}[t]
    \centering
    \begin{minipage}{0.4\textwidth}
        \centering
        \includegraphics[width=\linewidth]{figures/evaluations/revision/dis_ratio_legend_v2.pdf}
        \vspace{-5mm}
    \end{minipage}

    \begin{subfigure}{0.48\linewidth}
        \includegraphics[width=\linewidth]{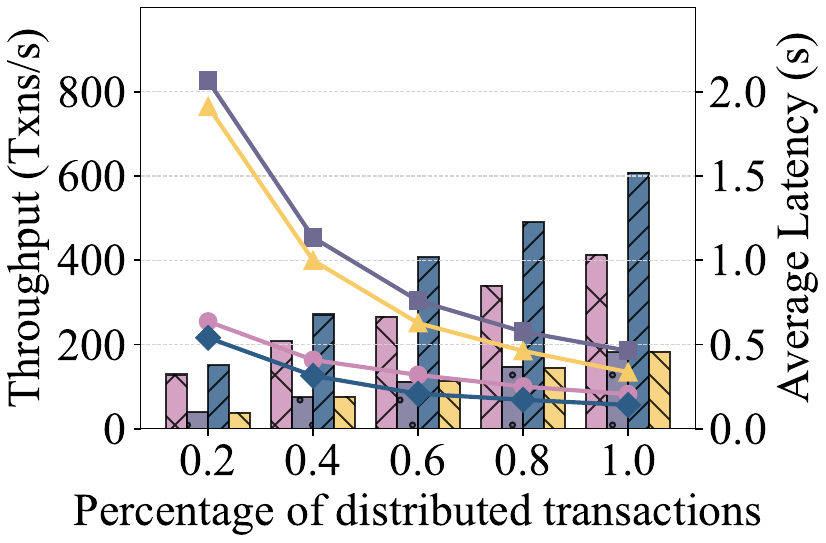}
        \vspace{-6mm}
        \caption{Payment}
        \label{Fig.tpcc.dis.payment}
    \end{subfigure}
    \begin{subfigure}{0.48\linewidth}
        \includegraphics[width=\linewidth]{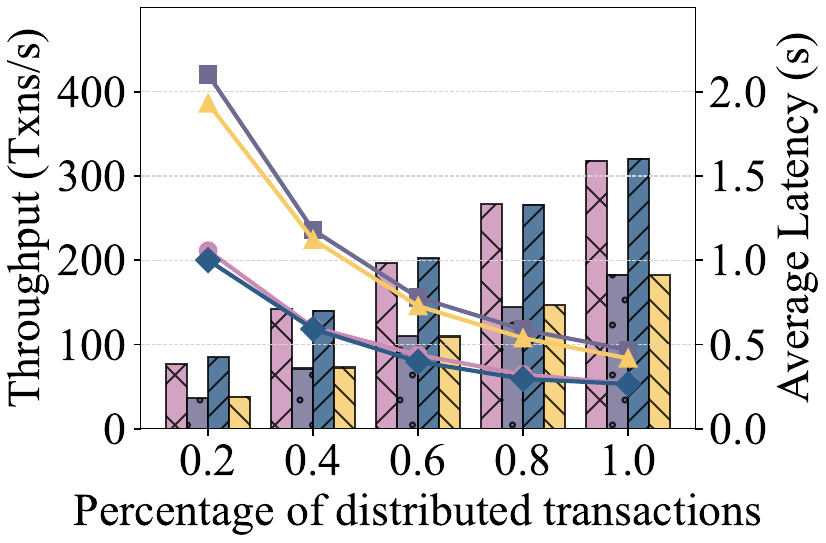}
        \vspace{-6mm}
        \caption{Neworder}
        \label{Fig.tpcc.dis.neworder}
    \end{subfigure}
    \vspace{-2mm}
    \caption{Impact of distributed transactions over TPC-C}
    \vspace{-5mm}
    \label{Fig.experiment.tpcc.dis}
\end{figure}

\begin{figure}[t]
    \centering
    \begin{minipage}{0.4\textwidth}
        \centering
        \includegraphics[width=\linewidth]{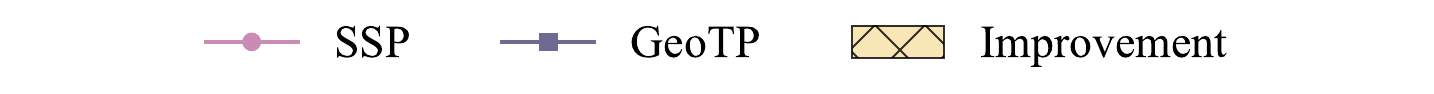}
        \vspace{-5mm}
    \end{minipage}
    \begin{subfigure}{0.48\linewidth}
        \includegraphics[width=\linewidth]{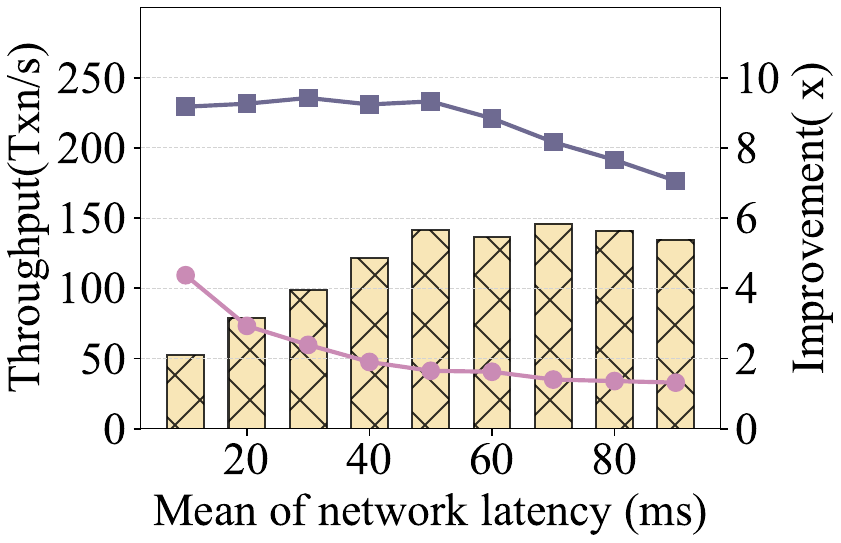}
        \vspace{-6mm}
        \caption{Fixed STD}
        \label{Fig.var.mean}
    \end{subfigure}
    \begin{subfigure}{0.48\linewidth}
        \includegraphics[width=\linewidth]{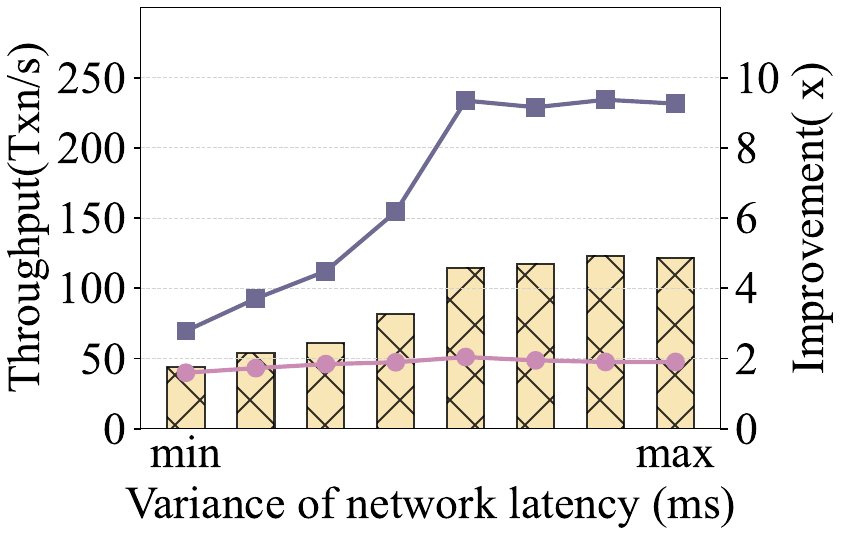}
        \vspace{-6mm}
        \caption{Fixed mean}
        \label{Fig.var.var}
    \end{subfigure}
    \vspace{-2mm}
    \caption{Impact of network latency configurations over YCSB}
    \label{Fig.experiment.var}
    \vspace{-5mm}
\end{figure}

\vspace{0.5ex} \noindent \textbf{TPC-C}: 
We control the ratio of distributed transactions in Payment and NewOrder by generating warehouseIDs and itemIDs in different data nodes. 
Figure~\ref{Fig.experiment.tpcc.dis} demonstrate that 
\dbname achieves 2.81x (2.04x) higher throughput and a 66.6\% (53\%) reduction in latency for Payment (NewOrder) transactions. It slightly outperforms Chiller due to the relatively low contention levels in our TPC-C workloads. Compared to other baselines, the performance improvement for NewOrder transactions is less significant than for Payment transactions, as Payment transactions tend to incur more contention.
This observation indicates the effectiveness of \dbname's heuristic
optimization for high-contention workloads. 

\subsection{Impact of Dynamic Network Latency}\label{sec:evaluation_dynamic_network}
In this subsection, we evaluate \dbname using YCSB under various network configurations, simulating by adjusting the network latency between the DM and data sources.

\noindent \textbf{Mean and standard deviation:}  
Figure~\ref{Fig.experiment.var} shows the result by varying mean and variance of network latency.
For example, when the mean latency is set to 20ms, the latencies between the middleware and three data nodes are 10ms, 20ms, and 30ms, respectively.
First, by fixing the standard deviation of network latency from DM to data sources, as the average latency increases, both \dbname and SSP throughputs decrease, but the relative advantage of \dbname over SSP becomes more pronounced. 
Then, by fixing the network mean between nodes, as the standard deviation increases, SSP performance remains relatively unchanged, while \dbname's performance continues to improve. 
This indicates that if only a few machines experience occasional latency spikes, it has a significant and severe impact on SSP but has minimal impact on \dbname. 

\begin{figure}[t]
    \centering
    \begin{subfigure}{0.45\linewidth}
        \includegraphics[width=\linewidth]{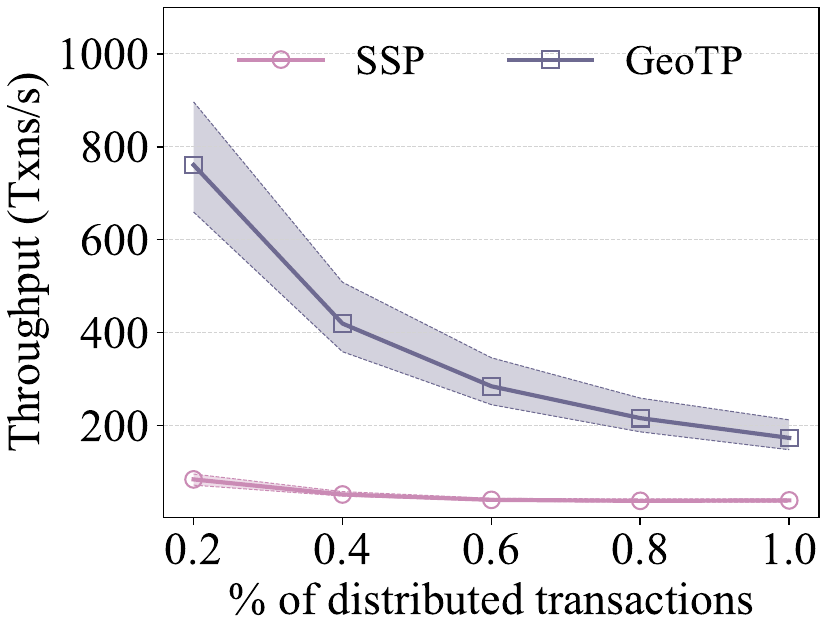}
        \vspace{-6mm}
        \caption{Random latency}
        \label{Fig.ycsb.rand.per}
    \end{subfigure}
    \begin{subfigure}{0.45\linewidth}
        \includegraphics[width=\linewidth]{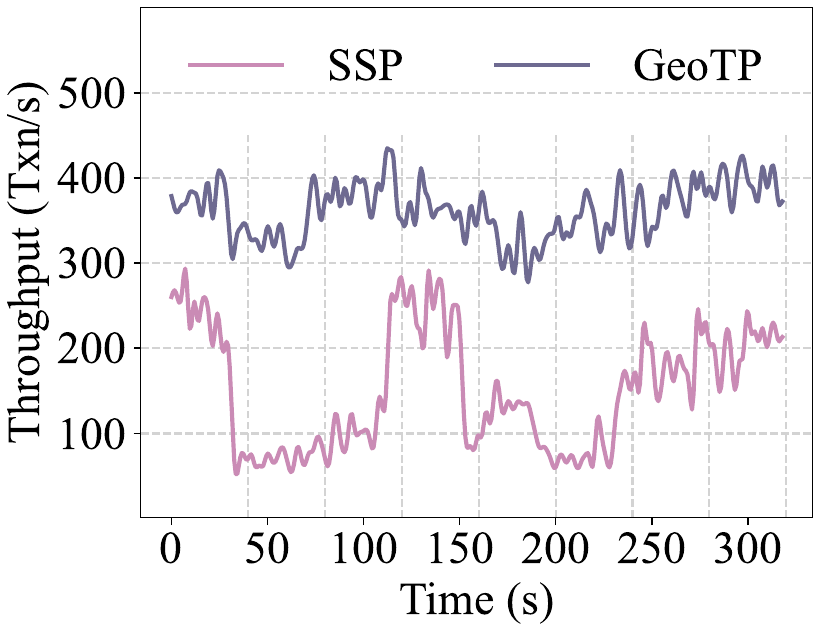}
        \vspace{-6mm}
        \caption{Dynamic latency}
        \label{Fig.ycsb.rand.timeline}
    \end{subfigure}
    \vspace{-2mm}
    \caption{Impact of random network latency over YCSB}
    \label{Fig.experiment.ycsb.rand}
    \vspace{-5mm}
\end{figure}

\noindent \textbf{Random latency:} 
We conduct experiments to evaluate \dbname with random network latencies by YCSB. As shown in Figure~\ref{Fig.ycsb.rand.per}, the solid line represents the average performance of running the experiment 20 times, while the portions filled with the shadow indicate the performance variations under different network environments. \dbname outperforms SSP in all scenarios with distributed transaction ratios ranging from 0.2 to 1.0. Further, when the network latency randomly fluctuates by a factor of 1.5 for some nodes, the performance jitter remains within 22.5\%. In the medium-contention workload, the average performance gains range from 4.5x to 9.1x.

\noindent \textbf{Online adaptivity:} We evaluate \dbname with an online dynamic network, adjusting the network latency every 40 seconds over a 320-second period. In Figure~\ref{Fig.ycsb.rand.timeline}, \dbname outperforms SSP in all scenarios and exhibits less sensitivity to dynamic network environments compared to SSP.
This capability of \dbname is attributed to its real-time network monitoring and latency-aware scheduling. In \dbname, we utilize the exponential weighted moving average algorithm~\cite{kurose2010computer} when we update the network latency. This helps \dbname balance temporary impacts and changes in trends.
Over time, \dbname demonstrates performance improvements ranging from 1.1x to 10.5x.

\begin{figure}[t]
    \centering
    \begin{minipage}{0.46\textwidth}
        \centering
        \includegraphics[width=\linewidth]{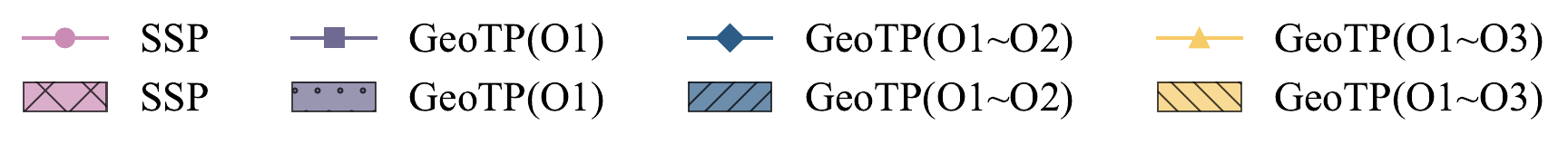}
        \label{Fig.experiment.ycsb.skew.legend}
        \vspace{-4mm}
    \end{minipage}

    \begin{minipage}{0.46\textwidth}
        \centering
        \includegraphics[width=\linewidth]{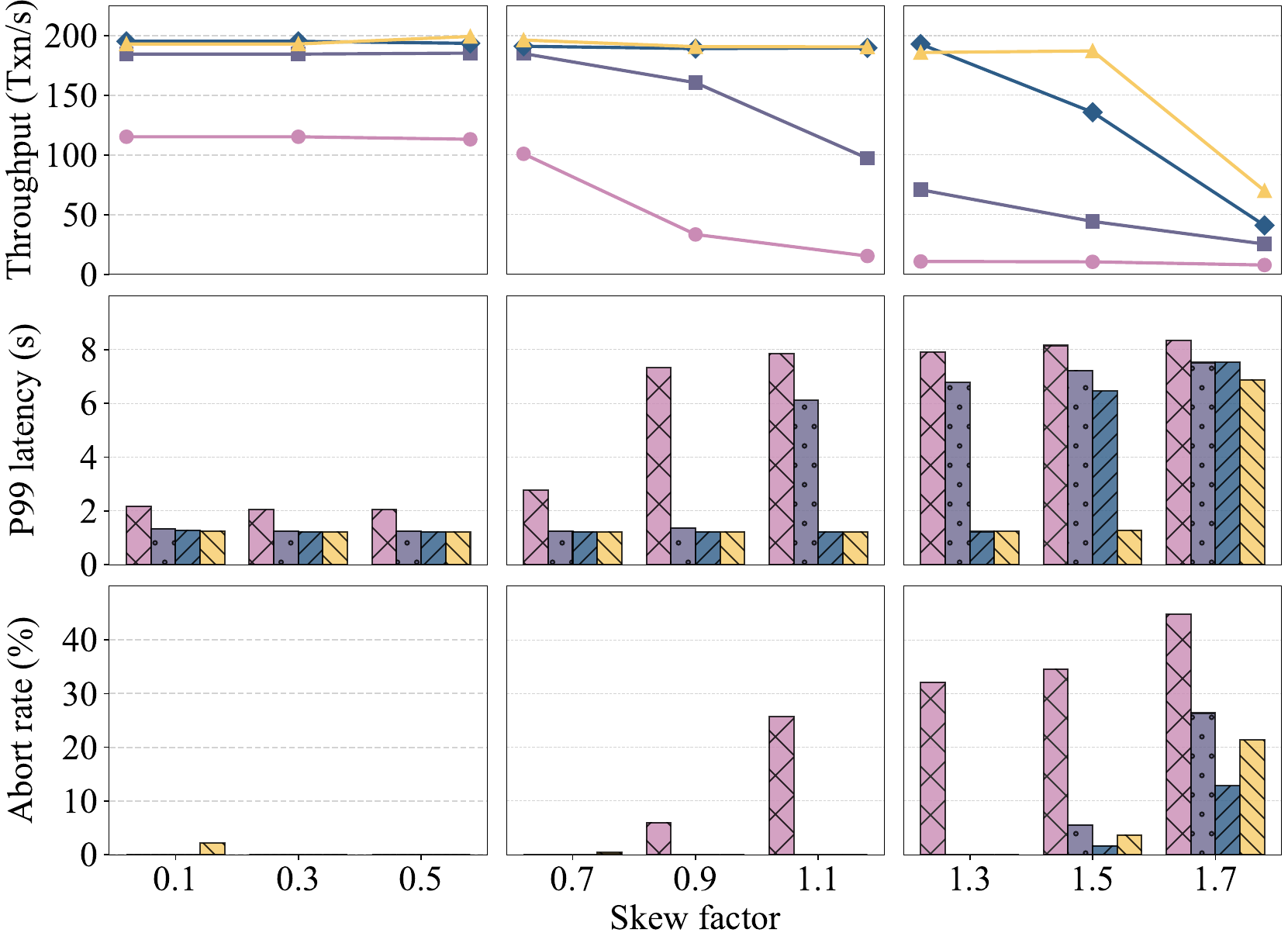}
        \vspace{-4mm}
    \end{minipage}
    \vspace{-2mm}
    \caption{Impact of optimizations over YCSB}
    \vspace{-6mm}
    \label{Fig.experiment.ycsb.skew}
\end{figure}

\subsection{Ablation Study}\label{sec:evaluation_optimization}
We now study the effectiveness of the three optimizations: 
(1) \textbf{O1}: the decentralized prepare mechanism 
as detailed in \S\ref{design-1},
(2) \textbf{O2}: the latency-aware scheduling mechanism 
as detailed in \S\ref{design-2},
and (3) \textbf{O3}: the high-contention workload optimization 
as elaborated in \S\ref{design-3}. 
Then, we use both O1 and O2 in \textbf{\dbname(O1 $\sim$ O2)}. Similarly in \textbf{\dbname(O1 $\sim$ O3)}, we use O1, O2 and O3. 
We compare \dbname and SSP with 50\% distributed transactions and a variety of skew factors (theta) as shown in Figure \ref{Fig.experiment.ycsb.skew}. 
The x-axis is partitioned into three segments denoting low (theta: 0.1 $\sim$ 0.5), medium (theta: 0.7 $\sim$ 1.1), and high (theta: 1.3 $\sim$ 1.7) contention scenarios. On the other hand, the y-axis illustrates throughput, p99 latency, and abort rate. 
The results demonstrate that \dbname achieves significantly higher throughput, reaching up to 17.7x greater than SSP. 
Meanwhile, \dbname reduces the abort rate by up to 32.1\% and p99 latency up to 84.3\% when compared to SSP.

\dbname outperforms SSP in all scenarios. However, the effectiveness of each optimization varies across different contentions. In low-contention workloads, the performance gains from O1 $\sim$ O3 are not particularly advantageous compared to O1 alone. In this case, the execution latency of a transaction primarily consists of network latency. As the contention among transactions is low, the benefits of scheduling subtransactions are not significant. Meanwhile, the abort ratio and p99 latency are low for all approaches. 
In medium-contention workloads, both \dbname(O1) and SSP decline.
In this case, transactions exhibit more data dependencies, and a transaction's execution latency comprises both network latency and local execution latency. O2 reduces the contention span and improves the concurrency. SSP's abort rate rises due to prolonged blocking time, resulting in lock wait timeouts. 
In high-contention workloads, the performance of all methods declines significantly due to critical lock waits, highlighting the insufficiency to consider only network latency when scheduling transactions. However, the degradation of \dbname (O1$\sim$O3) is minimal, with its p99 latency remaining the lowest, due to O3's ability to partially incorporate execution latency into scheduling while restricting access to hot records. The abort rate for \dbname (O1$\sim$O3) is lower than both SSP and \dbname (O1) but slightly higher than \dbname (O1$\sim$O2) because O3 mitigates lock contention by selectively blocking or aborting transactions.

\subsection{Impact of Heterogeneous Databases}
\label{sec.eva.heterogeneous}
\begin{table}[]
\footnotesize
\caption{Impact of heterogeneous deployments with various distributed transaction ratios (dr) over YCSB – \textnormal{Throughput (Txn/s) and Average latencies (ms)}}
\resizebox{\linewidth}{!}{
\begin{tabular}{ll|cc|cc}
    \toprule
    \multicolumn{2}{c|}{}                & \multicolumn{2}{c|}{\textbf{dr=25\%}} & \multicolumn{2}{c}{\textbf{dr=75\%}} \\
    \multicolumn{2}{c|}{}                & Throughput & Average latency & Throughput & Average latency \\
    \midrule  
    \multirow{2}{*}{\textbf{S1}}    
    & \textbf{SSP}       & 58.7 & 1441.8 & 33.3 & 1815.8 \\
    & \textbf{\dbname}   & 437.8 & 176.0 & 123.5 & 689.6 \\
    \multirow{2}{*}{\textbf{S2}}    
    & \textbf{SSP}       & 74.0 & 1069.9 & 35.5 & 2192.9 \\
    & \textbf{\dbname}   & 340.7 & 220.8 & 131.8 & 650.1 \\
    \multirow{2}{*}{\textbf{S3}}     
    & \textbf{SSP}       & 70.3 & 901.8 & 25.2 & 2112.6 \\
    & \textbf{\dbname}   & 425.5 & 198.8 & 116.6 & 632.3 \\
    \bottomrule
\end{tabular}
}
\label{Tbl.experiment.heterogeneous}
\vspace{-5mm}
\end{table}

We now evaluate the performance of \dbname when deployed on heterogeneous data sources, i.e., either MySQL or PostgreSQL. 
We denote those data nodes as {\small $N_1\sim N_4$}, respectively. 
We consider three scenarios: (\textbf{S1}) MySQL is deployed on nodes {\small $N_1 \sim N_4$}; (\textbf{S2}) PostgreSQL is deployed on nodes {\small $N_1$ \& $N_3$}, and MySQL is deployed on nodes {\small $N_2$ \& $N_4$}; (\textbf{S3}) PostgreSQL is deployed on nodes {\small $N_1 \sim N_4$}. 
As observed in Table~\ref{Tbl.experiment.heterogeneous}, \dbname outperforms baselines in all deployments. 
The throughput improvement ranges from 3.6x to 7.5x, with the average latency reduction varying between 62\% and 87.8\%. MySQL and PostgreSQL suffer from the long lock contention span in geo-distributed scenarios, and \dbname can improve the performance in these deployments.

\subsection{Comparison with YugabyteDB}\label{sec:evaluation_yugabyte}

\begin{figure}[t]
    \centering
    \begin{subfigure}{0.45\linewidth}
        \includegraphics[width=\linewidth]{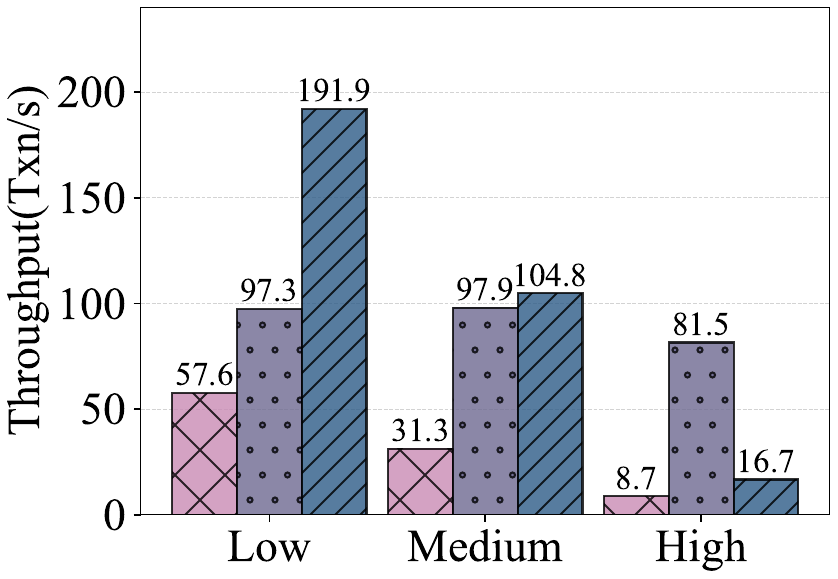}
        \vspace{-6mm}
        \caption{Throughput}
        \label{Fig.ygdb.per}
    \end{subfigure}
    \begin{subfigure}{0.45\linewidth}
        \includegraphics[width=\linewidth]{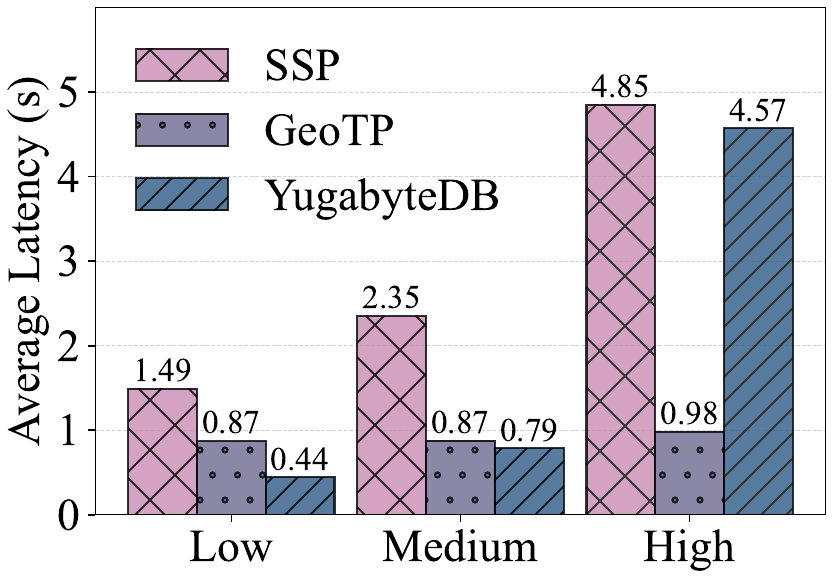}
        \vspace{-6mm}
        \caption{Average latency}
        \label{Fig.ygdb.lat}
    \end{subfigure}
    \vspace{-1mm}
    \caption{Comparison with YugabyteDB over YCSB}
    \label{Fig.experiment.ygdb}
    \vspace{-5mm}
\end{figure}

We compare the performance of \dbname with YugabyteDB, an advanced distributed database.
To ensure a fair comparison, we deploy YugabyteDB across 4 data nodes and partition the data into these nodes.
We measure the throughput and average latency over YCSB with varying contention levels and plot the results in Figure~\ref{Fig.experiment.ygdb}.
\dbname achieves a 4.88x improvement in the high-contention workload due to the proposed latency-aware scheduling and heuristic optimization. 
In medium-contention workloads, \dbname is on par with YugabyteDB. 
However, YugabyteDB outperforms \dbname in low contention, due to its ability to perform data updates asynchronously for single-row/single-shard transactions after commitment.
While \dbname does not directly modify the underlying data source, thus typically lacking this optimization.
In low-contention workloads, where transaction contentions no longer dominate performance, the performance gap due to these fundamental codebase differences of \dbname and YugabyteDB becomes more apparent. 
However, \dbname can benefit from the asynchronous update if supported by the underlying data source, making this technique orthogonal to our proposal.

\subsection{Impact of Transaction Length}
\label{sec:evaluation_length}

\begin{figure}[t]
    \centering
    \begin{subfigure}{0.3\linewidth}
        \includegraphics[width=\linewidth]{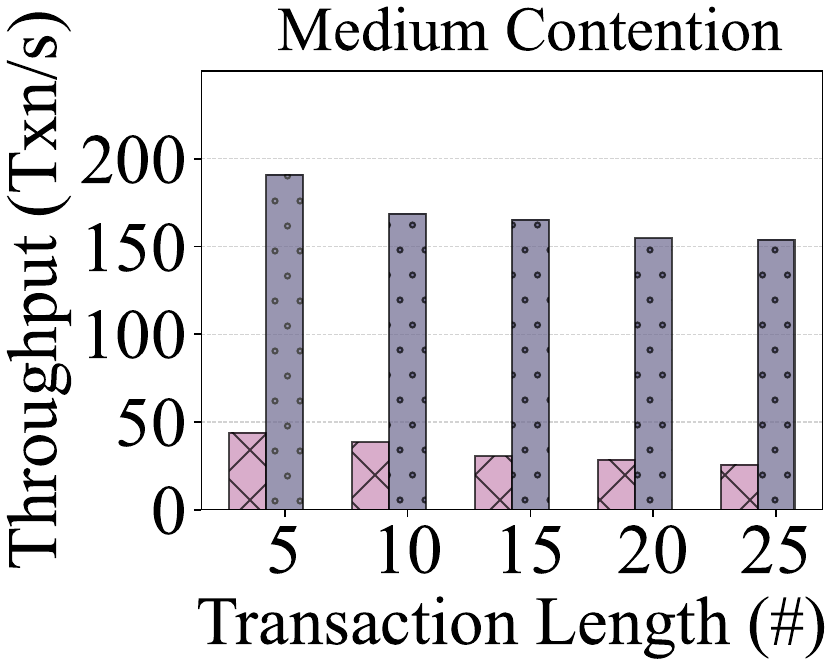}
        \vspace{-6mm}
        \caption{Length}
        \label{Fig.length.len}
    \end{subfigure}
    \begin{subfigure}{0.3\linewidth}
        \includegraphics[width=\linewidth]{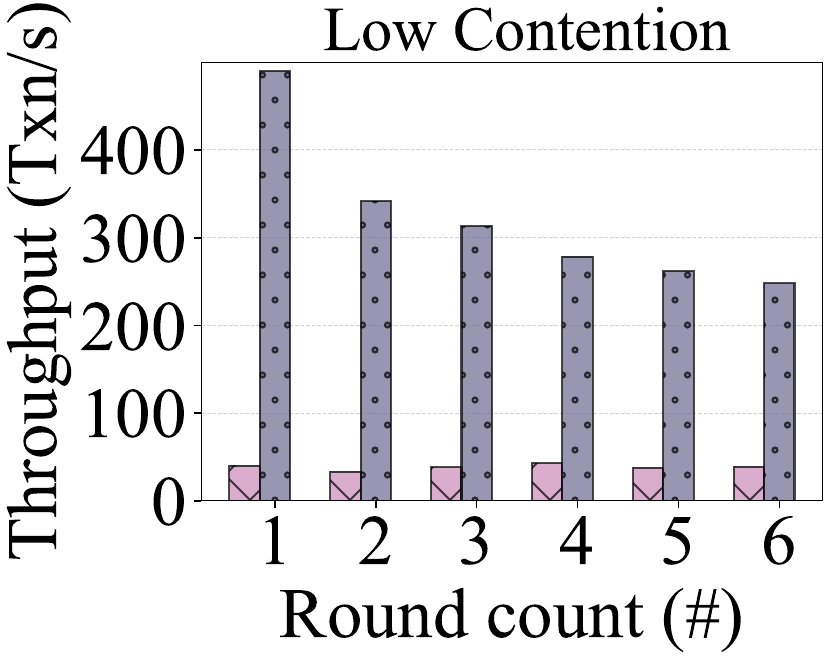}
        \vspace{-6mm}
        \caption{Round}
        \label{Fig.length.round.low}
    \end{subfigure}
    \begin{subfigure}{0.3\linewidth}
        \includegraphics[width=\linewidth]{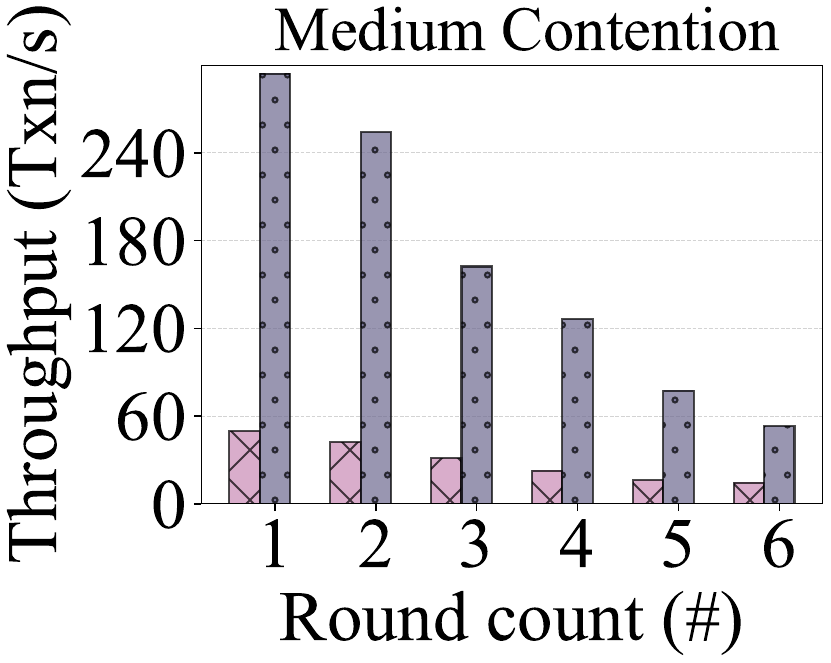}
        \vspace{-6mm}
        \caption{Round}
        \label{Fig.length.round.med}
    \end{subfigure}
    \vspace{-1mm}
    \caption{Impact of transaction length over YCSB}
    \label{Fig.experiment.length}
    \vspace{-6mm}
\end{figure}

We now study the impact of the transaction length and interactive rounds. We evaluate \dbname and baseline in medium contention workloads with 20\% distributed transaction. Figure~\ref{Fig.experiment.length} shows results for two settings. 
First, we examine the performance of fixed one-interaction round transactions while adjusting transaction length. As seen in Figure \ref{Fig.length.len}, throughput for both \dbname and SSP decreases by 19.1\% and 41.3\% as the length increases from 5 to 25. It remains relatively stable compared to the number of interaction rounds. Next, we vary the number of interaction rounds and evaluate \dbname in both low- and medium-contention workloads. As shown in Figure \ref{Fig.length.round.low} and~\ref{Fig.length.round.med}, with 6 interaction rounds, \dbname outperforms SSP by 1.5x in low-contention and 3.4x in medium-contention environments. This indicates that network round trip is the primary bottleneck. As the number of rounds increases, the advantages of the decentralized prepare mechanism decrease, while latency-aware scheduling and high-contention optimizations continue to provide performance gains.

\extended{
\subsection{Multi-region Deployment}
\dbname obeys the deployment of previous database middleware such as Shardingsphere, where a database middleware manages multiple underlying databases. Moreover, the architecture can scale to deploy multiple middleware close to clients in different regions, and the optimizations of \dbname do not rely on global centralized components. We have evaluated the performance of \dbname in different deployments, shown in Figure \ref{Fig.eva.multi_client}. There are two database middleware, denoted as $DM_1$ and $DM_2$, that connect to four data sources with distinct network latencies. For $DM_1$, the latencies to each data source are 0, 27 ms, 73 ms, and 251 ms, while for $DM_2$, they are 251 ms, 226 ms, 175 ms, and 0, respectively. Each DM is co-located with a client on the same server. 
\begin{figure}
    \centering
    \includegraphics[width=0.25\textwidth]{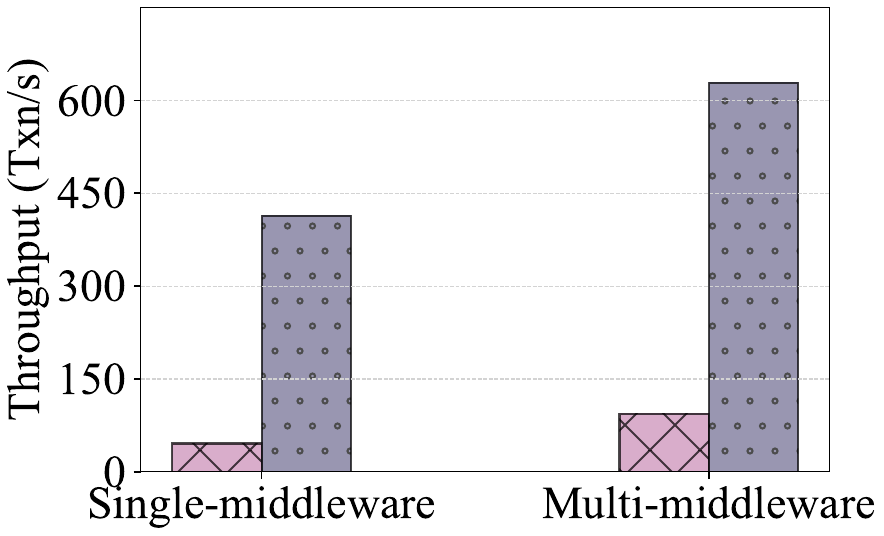}
    \caption{Clients in multiple regions over YCSB}
    \label{Fig.eva.multi_client}
    \vspace{-4mm}
\end{figure}
\dbname achieves up to 6.7x higher performance than SSP. 
The results demonstrate that \dbname can efficiently extend to multi-region deployment.
}

\section{Related Work\label{sec.related}}



\noindent\textbf{Database middleware techniques.} Substantial efforts have been dedicated to enhancing transactional capability across heterogeneous databases. For example, 
Skeena~\cite{DBLP:conf/sigmod/ZhangH0L22} efficiently integrates different engines within the same database system, and each engine operates autonomously. It identifies disparities in engine processing capabilities and employs a snapshot map in shared memory to ensure isolation. In contrast, Cherry Garica~\cite{DBLP:conf/icde/DeyFR15}, Epoxy~\cite{DBLP:journals/pvldb/Kraft0ZBSYZ23}, and ScalarDB~\cite{DBLP:journals/pvldb/YamadaSIN23} implement transaction management and concurrency control over the abstractions of underlying engines, making it an extendable to more kinds of engines, including NoSQLs. 
Other solutions focus on managing connections between databases and clients~\cite{ProxySQL, DBLP:journals/pvldb/ButrovichRRLZSP23}. These solutions enable the routing of statements to one or multiple database servers, thereby offering scalability and high performance. 
\dbname is designed for database middleware with geo-distributed data sources. These techniques are orthogonal to our contributions. 

\noindent\textbf{Other distributed transaction processing techniques.} 
Except for the studies~\cite{DBLP:conf/eurosys/ChenSJRLWZCC21, DBLP:journals/pvldb/YanC16, DBLP:conf/sigmod/ZamanianSBK20} in \S~\ref{design-1}, there are some works explore the scheduling and locking techniques in geo-distributed scenarios. Some approaches focus on reducing network round trips in conventional networks \cite{DBLP:journals/pvldb/MaiyyaNAA19, DBLP:journals/pvldb/ZhangLZXLXHYD23, DBLP:conf/sosp/ZhangSSKP15,DBLP:conf/icde/ZhengZLYCPD24_Lion, DBLP:conf/usenix/CowlingL12}. For instance, Multi-level 2PC~\cite{DBLP:journals/tods/MohanLO86} reduces costly WAN communication by organizing participants hierarchically, though it incurs higher LAN coordination overhead. 
Another line of research aims to reduce lock contention by enforcing partial or full determinism in concurrency control. 
Calvin \cite{DBLP:conf/sigmod/ThomsonDWRSA12} and Detock \cite{DBLP:journals/pacmmod/NguyenMA23} use a global agreement scheme to sequence lock requests deterministically. 
Deterministic techniques require a priori knowledge of read-set and write-set. Moreover, the methods mentioned above involve significant modifications to the database system or kernel-level transaction protocol, which limits their applicability in database middleware. In contrast, \dbname is a lightweight approach that reduces WAN communication cost and locks contention by accounting for differential network latency—an aspect often overlooked by the above approaches. 

\section{Conclusion\label{sec.conclusion}}
In this paper, we present \dbname, a latency-aware geo-distributed transaction processing in database middlewares 
without modifying the database kernels. 
The core idea of \dbname is to minimize latency and reduce the lock contention span of distributed transactions.
To achieve this, we introduce a decentralized prepare mechanism, which eliminates one WAN round trip for each distributed transaction.
Furthermore, we present a latency-aware scheduling approach that postpones the lock acquisition time for some subtransactions.
Lastly, we enhance latency-aware scheduling with heuristic optimizations for high-contention workloads.
Extensive experiments on YCSB and TPC-C show that \dbname outperforms baselines. Compared to other distributed databases, \dbname performs comparably in medium-contention workloads and excels in high-contention workloads.


\balance

\bibliographystyle{IEEEtran}
\bibliography{sample}


\end{document}